\documentclass[twocolumn]{aastex62}
\newcommand{\Gaia}{{\sl Gaia}}

\newcommand{\WISE}{{\sl WISE}}

\newcommand{\Lsun}{\mbox{$L_{\sun}$}}

\newcommand{\Mjup}{\mbox{$M_{\rm Jup}$}}
\newcommand{\Rjup}{\mbox{$R_{\rm Jup}$}}
\newcommand{\degree}{\mbox{$^{\circ}$}}
\newcommand{\perpix}{\mbox{pixel$^{-1}$}}

\newcommand{\kms}{\mbox{km\,s$^{-1}$}}
\newcommand{\masyr}{\hbox{mas\,yr$^{-1}$}}

\newcommand{\Ks}{\mbox{$K_S$}}

\newcommand{\CHs}{\mbox{$CH_4s$}}
\newcommand{\Kn}{\mbox{$K_{\rm H2}$}}
\newcommand{\Mtot}{\mbox{$M_{\rm tot}$}}

\newcommand{\Lbol}{\mbox{$L_{\rm bol}$}}
\newcommand{\fbol}{\mbox{$f_{\rm bol}$}}

\newcommand{\Teff}{\mbox{$T_{\rm eff}$}}
\newcommand{\logg}{\mbox{$\log(g)$}}

\newcommand{\Lp}{\mbox{${L^\prime}$}}

\newcommand{\fldg}{\textsc{fld-g}}

\newcommand{\objlong}{WISE~J072003.20$-$084651.2}
\newcommand{\obj}{WISE~J0720$-$0846}

\newcommand{\etal}{et~al.}

\begin{document}

\title{
WISE~J072003.20$-$084651.2B Is A Massive T~Dwarf\footnote{
Data presented herein were obtained at the W.M.\ Keck Observatory, which is operated as a scientific partnership among the California Institute of Technology, the University of California, and the National Aeronautics and Space Administration. The Observatory was made possible by the generous financial support of the W.M.\ Keck Foundation.
}\footnote{
Based on data obtained with WIRCam, a joint project of CFHT, Taiwan, Korea, Canada, France, at the Canada-France-Hawaii Telescope, which is operated by the National Research Council of Canada, the Institute National des Sciences de l'Univers of the Centre National de la Recherche Scientifique of France, and the University of Hawaii.
}}

\author{Trent J.\ Dupuy}
\affiliation{Gemini Observatory, Northern Operations Center, 670 N.\ A'ohoku Place, Hilo, HI 96720, USA}
\author{Michael C.\ Liu}
\affiliation{Institute for Astronomy, University of Hawaii, 2680 Woodlawn Drive, Honolulu, HI 96822, USA}
\author{William M.\ J.\ Best}
\affiliation{The University of Texas at Austin, Department of Astronomy, 2515 Speedway C1400, Austin, TX 78712, USA}
\author{Andrew W.\ Mann}
\affiliation{Department of Physics and Astronomy, University of North Carolina at Chapel Hill, Chapel Hill, NC 27599-3255, USA}
\author{Michael A.\ Tucker}
\affiliation{Institute for Astronomy, University of Hawaii, 2680 Woodlawn Drive, Honolulu, HI 96822, USA}
\author{Zhoujian Zhang}
\affiliation{Institute for Astronomy, University of Hawaii, 2680 Woodlawn Drive, Honolulu, HI 96822, USA}
\author{Isabelle Baraffe}
\affiliation{University of Exeter, Physics and Astronomy, EX4 4QL Exeter, UK}
\affiliation{\`Ecole Normale Sup\'erieure, Lyon, CRAL (UMR CNRS 5574), Universit\'e de Lyon, France}
\author{Gilles Chabrier}
\affiliation{\`Ecole Normale Sup\'erieure, Lyon, CRAL (UMR CNRS 5574), Universit\'e de Lyon, France}
\affiliation{University of Exeter, Physics and Astronomy, EX4 4QL Exeter, UK}
\author{Thierry Forveille}
\affiliation{Universit\'e Grenoble Alpes, CNRS, IPAG, F-38000 Grenoble, France}
\author{Stanimir A. Metchev}
\affiliation{The University of Western Ontario, Department of Physics and Astronomy, 1151 Richmond Avenue, London, ON N6A 3K7, Canada}
\author{Pascal Tremblin}
\affiliation{Maison de la Simulation, CEA, CNRS, Univ. Paris-Sud, UVSQ, Universit\'e Paris-Saclay, F-91191 Gif-sur-Yvette, France}
\author{Aaron Do}
\affiliation{Institute for Astronomy, University of Hawaii, 2680 Woodlawn Drive, Honolulu, HI 96822, USA}
\author{Anna V.\ Payne}
\affiliation{Institute for Astronomy, University of Hawaii, 2680 Woodlawn Drive, Honolulu, HI 96822, USA}
\author{B.\ J.\ Shappee}
\affiliation{Institute for Astronomy, University of Hawaii, 2680 Woodlawn Drive, Honolulu, HI 96822, USA}
\author{Charlotte Z.\ Bond}
\affiliation{Institute for Astronomy, University of Hawaii, 2680 Woodlawn Drive, Honolulu, HI 96822, USA}
\author{Sylvain Cetre}
\affiliation{W.\ M.\ Keck Observatory, 65-1120 Mamalahoa Hwy, Kamuela, HI, USA}
\author{Mark Chun}
\affiliation{Institute for Astronomy, University of Hawaii, 2680 Woodlawn Drive, Honolulu, HI 96822, USA}
\author{Jacques-Robert Delorme}
\affiliation{California Institute of Technology, Department of Astronomy, 1200 E. California Blvd., Pasadena, CA 91125, USA}
\author{Nemanja Jovanovic}
\affiliation{California Institute of Technology, Department of Astronomy, 1200 E. California Blvd., Pasadena, CA 91125, USA}
\author{Scott Lilley}
\affiliation{W.\ M.\ Keck Observatory, 65-1120 Mamalahoa Hwy, Kamuela, HI, USA}
\author{Dimitri Mawet}
\affiliation{California Institute of Technology, Department of Astronomy, 1200 E. California Blvd., Pasadena, CA 91125, USA}
\author{Sam Ragland}
\affiliation{W.\ M.\ Keck Observatory, 65-1120 Mamalahoa Hwy, Kamuela, HI, USA}
\author{Ed Wetherell}
\affiliation{W.\ M.\ Keck Observatory, 65-1120 Mamalahoa Hwy, Kamuela, HI, USA}
\author{Peter Wizinowich}
\affiliation{W.\ M.\ Keck Observatory, 65-1120 Mamalahoa Hwy, Kamuela, HI, USA}

\begin{abstract}

\noindent We present individual dynamical masses for the nearby M9.5+T5.5 binary WISE~J072003.20$-$084651.2{AB}, a.k.a.\ Scholz's star. Combining high-precision CFHT/WIRCam photocenter astrometry and Keck adaptive optics resolved imaging, we measure the first high-quality parallactic distance ($6.80_{-0.06}^{+0.05}$\,pc) and orbit ($8.06_{-0.25}^{+0.24}$\,yr period) for this system composed of a low-mass star and brown dwarf. We find a moderately eccentric orbit ($e = 0.240_{-0.010}^{+0.009}$), incompatible with previous work based on less data, and dynamical masses of $99\pm6$\,\Mjup\ and $66\pm4$\,\Mjup\ for the two components. The primary mass is marginally inconsistent (2.1$\sigma$) with the empirical mass--magnitude--metallicity relation and models of main-sequence stars.  The relatively high mass of the cold ($\Teff = 1250\pm40$\,K) brown dwarf companion indicates an age older than a few Gyr, in accord with age estimates for the primary star, and is consistent with our recent estimate of $\approx$70\,\Mjup\ for the stellar/substellar boundary among the field population. Our improved parallax and proper motion, as well as an orbit-corrected system velocity, improve the accuracy of the system's close encounter with the solar system by an order of magnitude. \obj{AB} passed within $68.7\pm2.0$\,kAU of the Sun $80.5\pm0.7$ kyr ago, passing through the outer Oort cloud where comets can have stable orbits. 

\end{abstract}


\section{Introduction}

While the nearest stars in the solar neighborhood are largely known, such stars residing in the Galactic Plane may still remain to be discovered, in particular low-luminosity ones.  As part of a search for such objects using 2MASS and \WISE, \citet{2014A&A...561A.113S} identified \objlong\ (hereinafter \obj) as a low-latitude ($b=+2.3\degree$) ultracool dwarf, with an estimated spectral type of M$9\pm1$ and a parallactic distance of 6--8~pc based on multi-catalog photometry and astrometry.
Followup by \citet{2015A&A...574A..64I} derived a spectral type of L$0\pm1$ based on optical and near-IR spectroscopy and refined the parallax measurement, making it the third closest known L dwarf.  They detected weak H$\alpha$ emission but not \ion{Li}{1} absorption, consistent with an old age and a mass above the stellar/substellar boundary.
\citet{2015AJ....149..104B} measured an optical spectral type of M9.5, found the spectrum to be consistent with solar metallicity, and observed variability (1\%--2\% amplitude) and flares (4\%--8\% amplitude) in TRAPPIST photometry.  They estimated an age for the system of 0.5--5.0~Gyr based on its 3-d space motion indicating old-disk kinematics.  \citet{2015AJ....150..180B} detected radio emission and a radial acceleration from Shane/Hamilton and Keck/NIRSPEC  radial velocity (RV) monitoring. Aside from its proximity, \obj\ is notable for being the star with the closest known past approach to the solar system, within $52^{+23}_{-14}$\,kAU only $70^{+15}_{-10}$\,kya\footnote{thousand years ago} \citep{2015ApJ...800L..17M}, though other stars are likely to come even closer in the future (e.g., Gl~710 will pass within $\approx$20~kAU in $\approx$1~Myr; \citealp{2018A&A...609A...8B, 2018RNAAS...2b..30D}).

The binarity of \obj\ has a convoluted history.
\citet{2014A&A...561A.113S} speculated that the star could be a close binary system based on the discrepancy between his parallactic and photometric distance determinations, though the two quantities were formally consistent given the large uncertainties.
\citet{2015A&A...574A..64I} found no convincing evidence for binarity based on the integrated-light spectrum and a refined parallax, though they noted a possible indication of radial velocity variability (at the 2$\sigma$ level) between measurements separated by three days.
\citet{2015AJ....149..104B} suggested that the system harbors a T5-dwarf companion, based on visual identification of peculiarities in the $H$-band integrated-light spectrum consistent with weak methane absorption.  (Quantitative analysis of the same spectrum by \citealp{2014ApJ...794..143B} did not flag this object as a candidate binary.)  Possible confirmation of such a companion came from their Keck adaptive optics (AO) imaging showing a candidate source at 0.14\arcsec\ separation, though our analysis here shows in fact this was a spurious detection.\footnote{Deacon \etal\ (2017) reported a possible detection of binarity from analyzing the ellipticity of seeing-limited Pan-STARRS~1 images obtained during $\approx$2010--2014, though they did not consider the result to be reliable.}  Follow-up AO imaging by \citet{2015AJ....150..180B} clearly showed a companion, with resolved near-IR photometry leading to a revised near-IR spectral type of T$5.5\pm0.5$ for the secondary.  They also used their resolved imaging and integrated-light radial velocities to estimate an orbital period of 4.1$^{+2.7}_{-1.3}$\,yr (c.f., our determination here of $8.06_{-0.25}^{+0.24}$\,yr).

As part of our ongoing effort to measure dynamical masses for ultracool dwarfs \citep[e.g.,][]{2008ApJ...689..436L, 2017ApJS..231...15D}, we have monitored \obj\ with Keck AO and CFHT wide-field astrometry for the past several years. There is only a handful of stars with T~dwarf companions at small separations amenable to orbit monitoring and thus dynamical mass determinations. \obj\ offers the valuable opportunity to compare inferences about stellar (late-M~dwarf) and substellar (T~dwarf) properties that depend on age or composition, under the conservative assumption of the system being coeval and co-composition.  For instance, independent age measurements for the two components can be compared to assess the reliability of the age-determination methods, a.k.a., the ``isochrone test'' \citep[e.g.,][]{2010ApJ...722..311L}.  In addition, the system straddles the stellar/substellar mass boundary and thus provides an opportunity to help delineate this boundary among the field population.  An ancillary product from such work is a high-precision proper motion and distance for the system (as it is not in Gaia DR2), which helps to understand its past dynamical interaction with the solar system. 

\begin{figure*}
\centerline{
\includegraphics[height=1.25in]{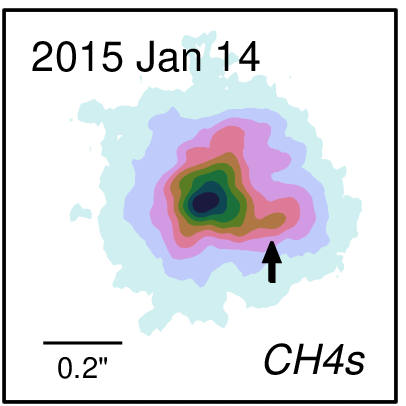}
\hfill
\includegraphics[height=1.25in]{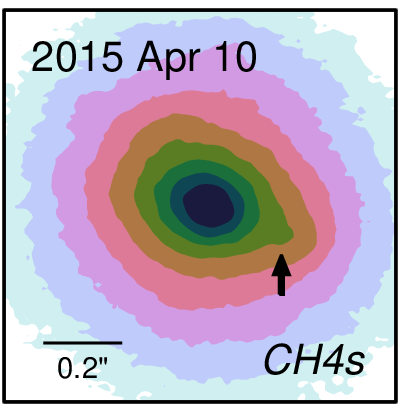}
\hfill
\includegraphics[height=1.25in]{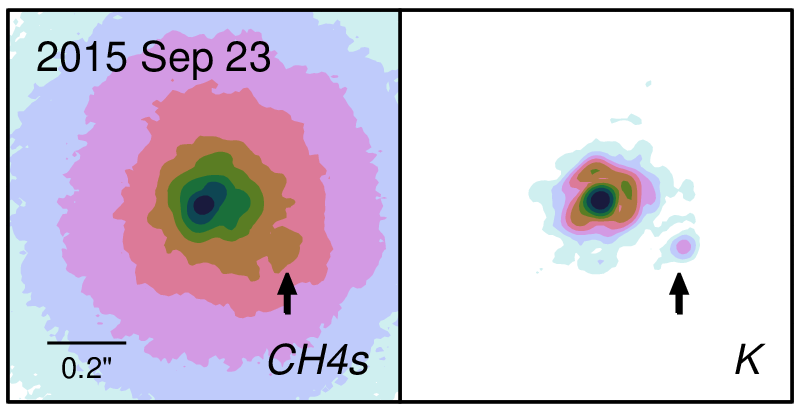}
\hfill
\includegraphics[height=1.25in]{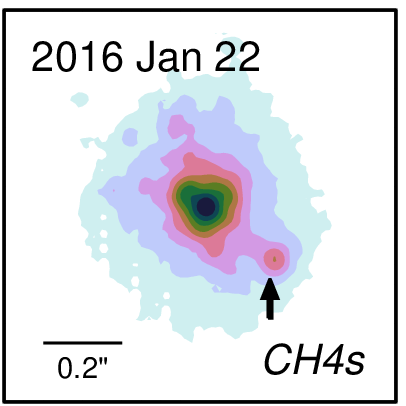}
}
\vskip 0.15in
\centerline{
\includegraphics[height=1.25in]{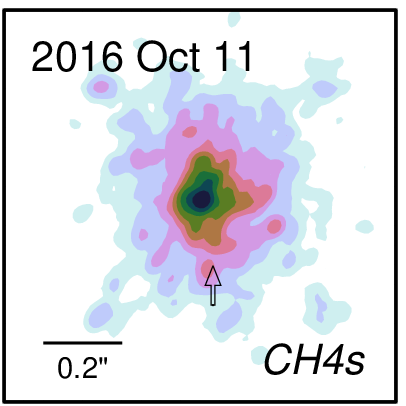}
\hfill
\includegraphics[height=1.25in]{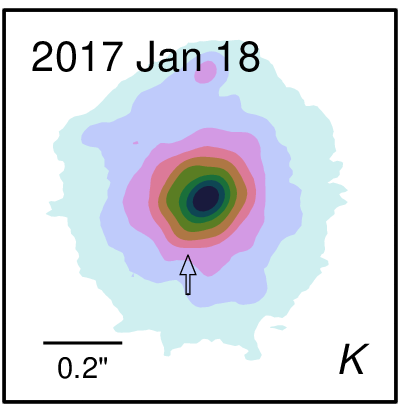}
\hfill
\includegraphics[height=1.25in]{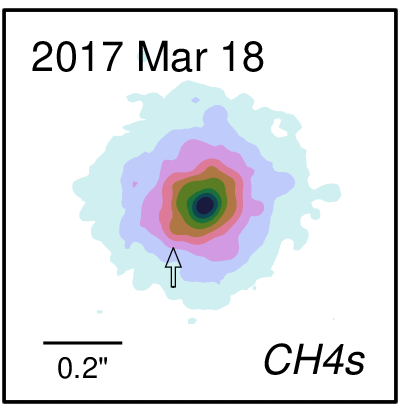}
\hfill
\includegraphics[height=1.25in]{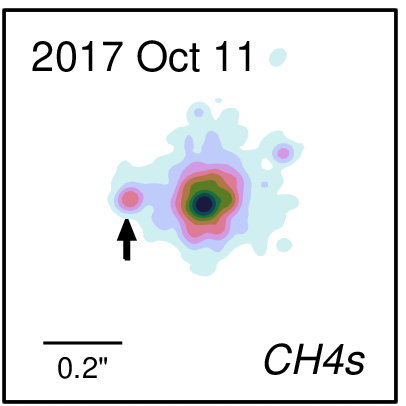}
\hfill
\includegraphics[height=1.25in]{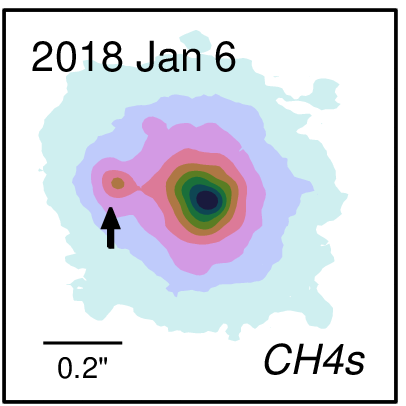}
}
\vskip 0.15in
\centerline{
\includegraphics[height=1.2in]{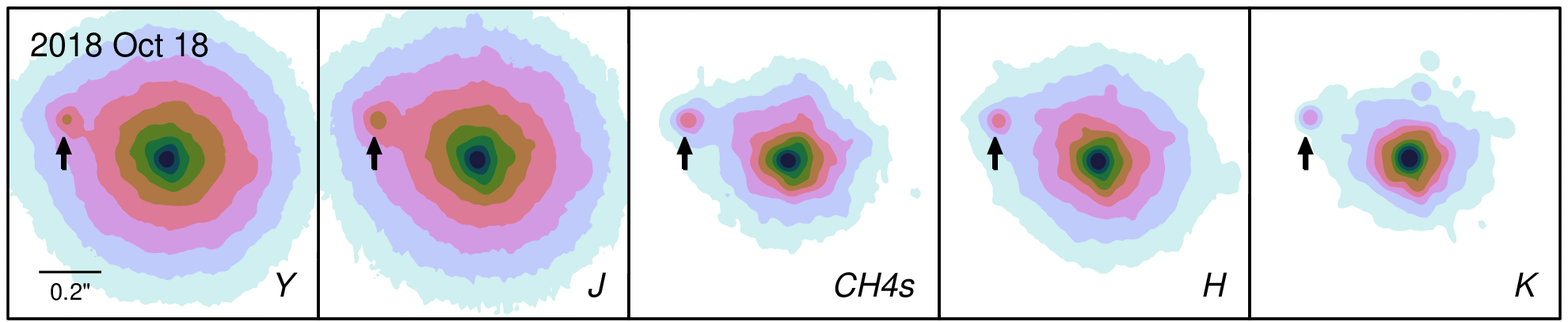}
}
\vskip 0.0 truein
\caption{Contour plots of typical individual exposures from our Keck LGS~AO data, with levels drawn in logarithmic intervals from unity down to 0.5\% of the peak flux in each image. The images are all 1\farcs0 across and have been rotated so that north is up. An arrow indicates the position of the companion at each epoch, where unfilled arrows indicate the predictions from our orbit fit at epochs where the companion is not resolved (2016~Oct~11~UT to 2017~Mar~18~UT). \label{fig:keck}}
\end{figure*}

\begin{deluxetable*}{lcccccccccc}
\tablecaption{Relative astrometry from Keck/NIRC2 Adaptive Optics Imaging \label{tbl:keck}}
\tablehead{
\colhead{Date} &
\colhead{Separation}    &
\colhead{PA}      &
\colhead{$\Delta{m}$} &
\colhead{Filter}  &
\colhead{$N_{\rm coadd}$}      &
\colhead{$N_{\rm exp}$}      &
\colhead{Airmass} &
\colhead{Strehl} &
\colhead{FWHM} &
\colhead{Notes}     \\[-4pt]
\colhead{(UT)}   &
\colhead{(mas)}    &
\colhead{(\degree)}      &
\colhead{(mag)} &
\colhead{} &
\colhead{$\times t_{\rm exp}$ (s)} &
\colhead{} &
\colhead{} &
\colhead{ratio} &
\colhead{(mas)} &
\colhead{} }
\startdata
2015~Jan~14 & $  188\pm5  $                   & $ 256.0\pm0.7  $                   & $ 2.87\pm0.10 $ & $\CHs$ &    $30\times1.0 $ & 11 & 1.14 & $ 0.084\pm0.015 $ &  $   72\pm6  $ & * \\
2015~Apr~10 & $  225\pm9  $                   & $ 252.6\pm1.5  $                   & $  3.9\pm0.5  $ & $\CHs$ &    $30\times1.0 $ &  9 & 1.45 & $ 0.025\pm0.005 $ &  $  143\pm17 $ & * \\
2015~Sep~23 & $  239\pm6  $                   & $ 238.2\pm1.1  $                   & $ 2.85\pm0.11 $ & $\CHs$ &    $50\times1.0 $ &  6 & 1.43 & $ 0.046\pm0.020 $ &  $   63\pm11 $ & * \\
2015~Sep~23 & $241.7\pm1.1$                   & $239.87\pm0.24 $                   & $3.880\pm0.020$ & $K   $ &    $10\times0.36$ &  2 & 1.46 & $ 0.326\pm0.029 $ &  $   53\pm0  $ & UX \\
2016~Jan~22 & $222.8\pm1.0$                   & $232.12\pm0.17 $                   & $ 3.30\pm0.15 $ & $\CHs$ &    $20\times1.0 $ & 11 & 1.42 & $ 0.110\pm0.021 $ &  $   53\pm3  $ & * \\
2017~Oct~11 & $188.5\pm0.8$                   & $ 85.77\pm0.06 $                   & $2.945\pm0.013$ & $\CHs$ &    $20\times1.0 $ & 10 & 1.19 & $ 0.197\pm0.008 $ &  $   46\pm2  $ & * \\
2018~Jan~6  & $232.8\pm1.1$                   & $ 79.32\pm0.06 $                   & $3.017\pm0.021$ & $\CHs$ &    $20\times1.0 $ & 11 & 1.14 & $ 0.081\pm0.014 $ &  $   60\pm4  $ & * \\
2018~Oct~18 & $346.1\pm1.4$                   & $67.937\pm0.022$                   & $2.910\pm0.009$ & $Y   $ &    $6\times20.0 $ &  3 & 1.25 & $ 0.077\pm0.005 $ &  $   61\pm2  $ & UX \\
2018~Oct~18 & $345.8\pm1.4$                   & $67.864\pm0.027$                   & $2.581\pm0.012$ & $J   $ &    $20\times1.0 $ &  4 & 1.27 & $ 0.030\pm0.006 $ &  $   61\pm5  $ & UX \\
2018~Oct~18 & $346.1\pm1.6$                   & $ 67.78\pm0.06 $                   & $2.971\pm0.008$ & $\CHs$ &    $20\times1.0 $ &  5 & 1.32 & $ 0.076\pm0.019 $ &  $   60\pm6  $ & * \\
2018~Oct~18 & $346.5\pm1.4$                   & $ 67.75\pm0.04 $                   & $3.201\pm0.019$ & $H   $ &    $20\times1.0 $ &  4 & 1.28 & $ 0.086\pm0.011 $ &  $ 57.8\pm1.5$ & UX \\
2018~Oct~18 & $346.4\pm1.4$                   & $ 67.61\pm0.03 $                   & $3.789\pm0.015$ & $K   $ &    $20\times1.0 $ &  8 & 1.29 & $ 0.214\pm0.025 $ &  $ 62.4\pm1.4$ & X  \\
2019~Apr~21 & $386.9\pm1.7$\tablenotemark{\#} & $ 63.44\pm0.09 $\tablenotemark{\#} & $3.867\pm0.013$ & $K   $ & $\phn3\times1.0 $ &  4 & 1.55 & $ 0.266\pm0.074 $ &  $   58\pm3  $ & PUX \\
2019~Apr~21 & $381.2\pm2.2$\tablenotemark{\#} & $ 63.53\pm0.27 $\tablenotemark{\#} & $ 3.01\pm0.03 $ & $\Lp $ &    $20\times0.2 $ & 43 & 1.61 & $ 0.687\pm0.055 $ &  $ 85.8\pm0.9$ & PX  
\enddata
\tablecomments{(*)~used in orbit fit; (P)~pyramid WFS observations; (U)~rms errors likely underestimated because $N_{\rm exp}\leq4$; (X)~not used in orbit fit.}
\tablenotetext{\#}{Astrometric calibration for the pyramid WFS system used with NIRC2 is not yet measured. We use the \citet{2016PASP..128i5004S} solution to quote representative numbers for separation and PA, but these should not be used for science. The orbit fit we derive in Section~\ref{sec:orbit} predicts a separation of $386\pm4$\,mas and PA of $63.28\pm0.16$\degree\ at this epoch.}
\end{deluxetable*}

\section{Observations}

\subsection{Keck/NIRC2 LGS AO \label{sec:keck}}

We began monitoring \obj\ on 2015~January~14~UT with NIRC2 and the laser guide star adaptive optics (LGS AO) system at the Keck~II telescope \citep{2004SPIE.5490..321B, 2006PASP..118..297W, 2006PASP..118..310V}. On the first two epochs (2015~Jan~15 and 2015~Apr~10~UT) we used a nearby star (USNO-B1.0~0812-0137391) that seemed it would provide better AO correction, but it turned out to be 1.2\,mag fainter than expected ($R\approx17.6$\,mag). At all other epochs we used the science target itself as the tip-tilt reference star, even though its optical faintness can make acquisition challenging. However, the tip-tilt sensor is very red sensitive, so it detected counts equivalent to a star of $\approx$14.8\,mag in spite of this target's actual $R$-band magnitude of 16.9\,mag \citep{2003AJ....125..984M}. At all epochs, we obtained data using NIRC2's medium-band filter centered on the $H$-band flux peak seen in T~dwarfs ($\CHs$; $\lambda_C = 1.592$\,\micron\ and $\Delta\lambda = 0.126$\,\micron). This filter offered the best compromise between the quality of the AO correction (better at longer wavelengths) and the SNR of the companion (better at bluer NIR wavelengths). Observing in a medium-band filter also mitigates the influence of differential chromatic refraction (DCR) that would otherwise introduce systematic offsets in our relative astrometry given the very different spectra of the two components. At some epochs we also obtained data in standard Mauna Kea Observatories (MKO) filters \citep{2002PASP..114..169S, 2002PASP..114..180T} for the purposes of measuring relative photometry. 

Figure~\ref{fig:keck} shows images from every epoch of our monitoring observations. At our first epoch, the companion was to the southwest and moving outward to wider separations. By 2015~Sep~23~UT, the companion had already started moving inward. Then, by 2016~Oct~11~UT, the companion was not resolved even though the image quality was comparable to or better than previous data sets. As we see later in our analysis, these nondetections are consistent with our derived orbit. A year later, the companion was recovered for the first time to the northeast at a separation of $0\farcs19$, and it has subsequently moved outward to its widest separation yet of $0\farcs38$.

At the epochs where the companion is resolved, we measured binary parameters (separation, PA, flux ratio) using similar methods as in our previous work \citep[e.g.,][]{2008ApJ...689..436L, 2010ApJ...721.1725D}. We fit an analytic, three-component Gaussian model to each point source when they are spatially blended, and when they are better separated we perform PSF-fitting using StarFinder \citep{2000A&AS..147..335D}. We then convert the measured $(x,y)$ positions into sky coordinates using the same methods as described in \citet{2016ApJ...817...80D} and \citet{2017ApJS..231...15D}, with the only difference being that we reversed the sign of the PA offsets of $0\fdg252$ and $0\fdg262$ in the \citet{2010ApJ...725..331Y} and \citet{2016PASP..128i5004S} calibrations, respectively, as found by \citet{2018AJ....155..159B}. At several recent epochs where the binary separation is rather wide ($\gtrsim0\farcs3$), the $4\times10^{-4}$ uncertainty in the pixel scale of NIRC2 is the dominant error term for our separation measurements. In most other cases, the rms of our dithered measurements dominates.

Table~\ref{tbl:keck} lists all our derived binary parameters, where each quoted error is the rms of measurements from individual images. For separations and PAs, the systematic uncertainties in the astrometric calibration, e.g., 0.004\,mas\,\perpix\ and $0\fdg020$ for the \citet{2016PASP..128i5004S} calibration, have been added in quadrature to the rms values. To assess the accuracy of our relative photometry, we examined the best $\CHs$ data sets and found an rms of 0.04\,mag in $\Delta\CHs$ values. We adopt this as a systematic noise floor for all of our relative photometry, most likely due to the limitations of our PSF modeling given that variability at this amplitude in the $H$ band is relatively uncommon \citep[e.g.,][]{2015ApJ...799..154M}.

\subsection{Keck/NIRC2 PyWFS AO \label{sec:pywfs}}

In addition to observations made using the facility LGS AO system, we also obtained $K$- and \Lp-band imaging with the newly installed infrared pyramid wavefront sensor (PyWFS) on Keck~II. The PyWFS will be part of the Keck Planet Imager and Characterizer \citep{2018SPIE10703E..06M}, an instrument optimized for high-contrast observations of faint red objects. The design and laboratory testing of the PyWFS is detailed in \citet{2018SPIE10703E..1ZB}. The instrument is currently being commissioned, and the data presented here were taken on the first shared-risk science night (2019~Apr~21~UT). The wavefront sensing was done in $H$ band ($\lambda = 1.65$\,\micron), with a pyramid modulation of $3\lambda/D$, thus enabling high-Strehl ratio NGS AO observations of red objects that are faint in the optical. 

\begin{figure}
\centerline{\includegraphics[height=1.25in]{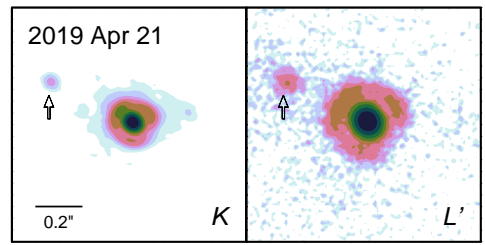}}
\caption{Contour plots of typical individual exposures from Keck/NIRC2 obtained with PyWFS AO. As in Figure~\ref{fig:keck}, levels are drawn in logarithmic intervals from unity to 0.5\% of the peak flux in each band, and the images are all 1\farcs0 across and have been rotated so that north is up. An arrow indicates the position of the companion, predicted from the orbit fit that does not use these data. \label{fig:pws}}
\end{figure}

Table~\ref{tbl:keck} includes astrometry measured from our PyWFS AO imaging, which is displayed in Figure~\ref{fig:pws}. The configuration of NIRC2 with the PyWFS differs from the facility AO configuration, as an additional dichroic sends $J$- and $H$-band light to the PyWFS, with the rest of the infrared light transmitted to NIRC2.  Precise astrometric calibration of NIRC2 in this configuration is a work in progress, so PyWFS astrometry should not yet be used for science. We note however that if we simply adopt the \citet{2016PASP..128i5004S} calibration, then the astrometry is in good agreement with our orbit determination from Section~\ref{sec:orbit} that predicts a separation of $386\pm4$\,mas and PA of $63.28\pm0.16$\degree\ at the PyWFS epoch.

\subsection{CFHT/WIRCam \label{sec:cfht}}

\begin{deluxetable*}{lccccccc}
\tablewidth{0pt}
\tablecaption{CFHT/WIRCam Astrometry of WISE~J0720$-$0846AB in Integrated Light \label{tbl:cfht}}
\tablehead{
\multicolumn{2}{c}{Observation Date} &
\colhead{R.A.}    &
\colhead{Decl.}    &
\colhead{$\sigma_{\rm R.A.}$} &
\colhead{$\sigma_{\rm Decl.}$} &
\colhead{Airmass}    &
\colhead{Seeing}     \\
\colhead{(UT)}   &
\colhead{(MJD)}  &
\colhead{(deg)}  &
\colhead{(deg)}  &
\colhead{(mas)}  &
\colhead{(mas)}  &
\colhead{}  &
\colhead{(arcsec)}}
\startdata
2015~Feb~8  & 57061.3162 & 110.01352251 & $-08.78093445$ &  3.5 &  5.2 & 1.175 & 0.60 \\
2015~Oct~22 & 57317.6506 & 110.01357698 & $-08.78094269$ &  5.6 &  5.7 & 1.139 & 0.58 \\
2015~Dec~23 & 57379.4871 & 110.01354626 & $-08.78095899$ &  4.1 &  4.5 & 1.140 & 0.71 \\
2016~Sep~12 & 57643.6386 & 110.01354930 & $-08.78096045$ &  7.3 &  9.6 & 1.613 & 0.58 \\
2017~Mar~15 & 57827.2925 & 110.01345537 & $-08.78099102$ &  2.4 &  4.2 & 1.170 & 0.57 \\
2017~Dec~5  & 58092.5417 & 110.01349091 & $-08.78104046$ &  3.3 &  7.2 & 1.142 & 0.67 \\
2018~Mar~2  & 58179.3278 & 110.01342954 & $-08.78104207$ &  3.8 &  6.2 & 1.168 & 0.74 \\
2018~Nov~1  & 58423.6050 & 110.01348236 & $-08.78107017$ &  4.3 &  4.1 & 1.149 & 0.64 \\
2018~Nov~21 & 58443.5650 & 110.01347304 & $-08.78107995$ &  1.5 &  2.5 & 1.140 & 0.62 \\
2019~Mar~23 & 58565.2709 & 110.01339713 & $-08.78107819$ &  2.2 &  2.9 & 1.168 & 0.51
\enddata
\tablecomments{The quoted uncertainties correspond to relative, not absolute, astrometric errors.}
\end{deluxetable*}

We first began monitoring \obj\ at the Canada-France-Hawaii Telescope (CFHT) on 2015~February~8~UT as part of the Hawaii Infrared Parallax Program \citep{2012ApJS..201...19D}, and since then obtained most of our data from our ongoing Large Program, the CFHT Infrared Parallax Program.  The facility infrared camera WIRCam \citep{2004SPIE.5492..978P} provides wide-field, seeing-limited imaging. For \obj, we obtained data using a narrow-band filter (0.032\,\micron\ bandwidth) in the $K$ band.  We refer to this filter as \Kn\ band because it is centered at 2.122\,\micron, the wavelength of the H$_2$~1-0~S(1) line. We use this filter for our brighter targets that risk saturation in wider bandpasses, and a side benefit of using such a narrow-band filter is that it renders DCR negligible. 

We typically obtained 18 images per epoch with exposure times of 5~s that resulted in SNR~$\approx$~600 on the target. We measured $(x,y)$ positions of the target and 145 reference stars in the field using SExtractor \citep[][using windowed Gaussian parameters]{1996A&AS..117..393B} and converted these to precision multi-epoch relative astrometry using a custom pipeline described in detail in our previous work \citep{2012ApJS..201...19D, 2016ApJ...833...96L}.  The absolute calibration of the linear terms of our astrometric solution was derived by matching low proper motion sources ($<30$\,\masyr) within the field of view to the 2MASS point source catalog \citep{2003tmc..book.....C}. The resulting astrometry for \obj\ in integrated light is given in Table~\ref{tbl:cfht}.  To convert our relative parallax and proper motion to an absolute frame, we used the mean and standard deviation of the simulated Galaxy population from the Besan\c{c}on model \citep{2003A&A...409..523R}.


\subsection{UH~2.2-m/SNIFS \label{sec:snifs}}

We obtained an optical spectrum of \obj\ with the SuperNova Integral Field Spectrograph \citep[SNIFS;][]{Aldering2002,Lantz2004} on the University of Hawai'i 2.2-m telescope on Maunakea on 2018~October~22~UT. SNIFS provides simultaneous coverage from 3200--9700\,\AA\ at a resolution of $R\simeq1200$, and the total exposure time of our observation was 3600~s. Details of our SNIFS reduction can be found in \citet{Bacon2001} and \citet{Gaidos2014}, which we briefly summarize here. The pipeline detailed in \citet{Bacon2001} performed dark, bias, and flat-field corrections, cleaned the data of bad pixels and cosmic rays, then fit and extracted the integral field unit spaxels into a one-dimensional (1D) spectrum. The \citet{Gaidos2014} reduction takes the 1D spectrum and performs flux calibration and telluric correction based on white dwarf standards taken throughout the night and a model of the atmosphere above Maunakea \citep{Buton2013}. 

\begin{figure*}
\centerline{
\includegraphics[width=6.5in]{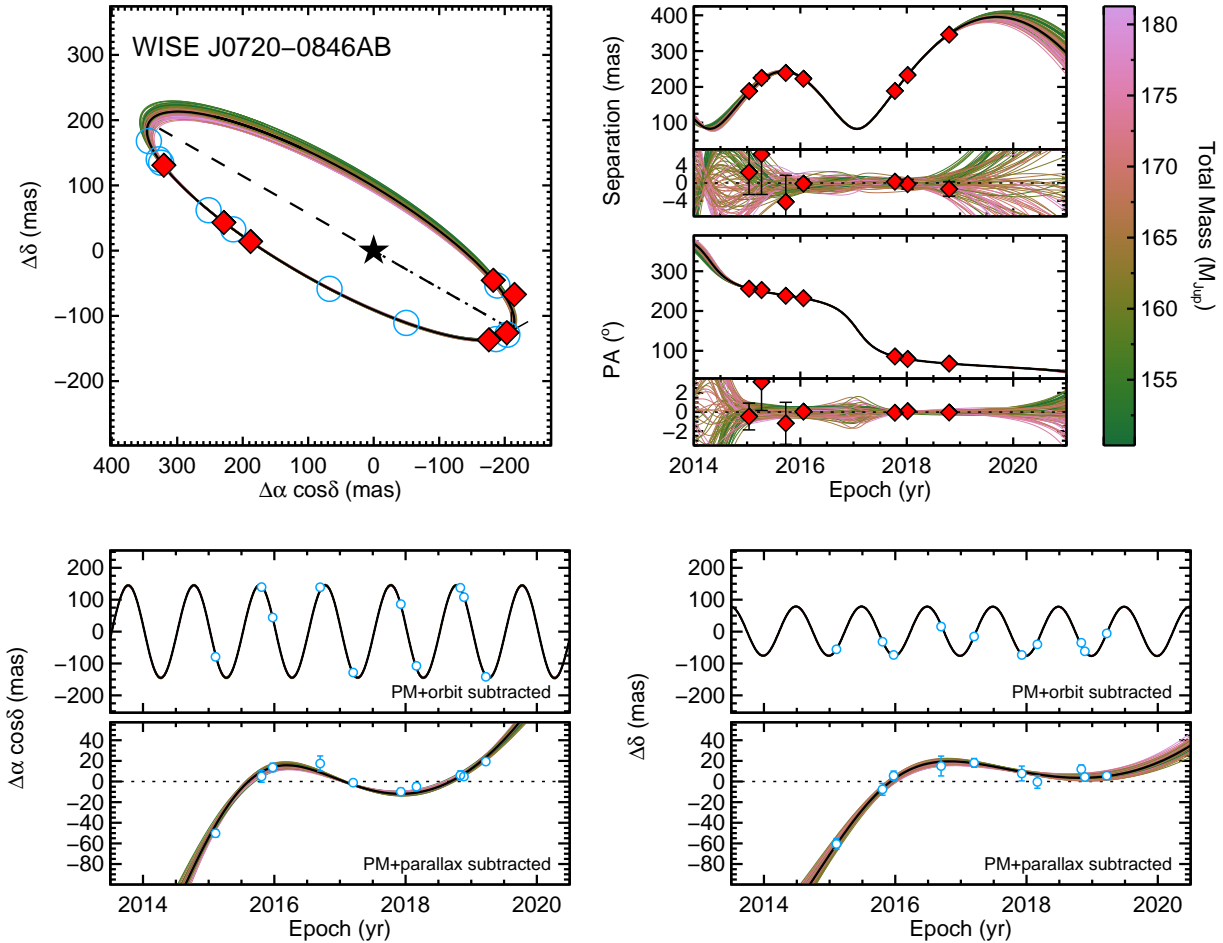}
}
\vskip 0.0 truein
\caption{Our astrometry and orbit determination for \obj{AB}. In all panels, the best-fit orbit is a thick black line, and 100 orbit solutions drawn randomly from our MCMC posterior are thin lines colored according to the dynamical total mass (color bar in top right panel). {\bf Top left:} Relative astrometry from Keck LGS AO imaging (red diamonds). The times corresponding to the observation epochs with CFHT/WIRCam are marked by open blue circles. The lines of nodes is indicated by a dashed line, and a dotted line connects the primary star to the point along the orbit corresponding to periastron passage (almost exactly overlapping with the SW node here). {\bf Top right:} Relative astrometry as a function of time with the lower subpanels showing residuals from the best-fit orbit. {\bf Bottom:} Integrated-light astrometry from CFHT/WIRCam as a function of time. Upper subpanels show the parallax curve that remains after subtracting proper motion and orbital motion (errors are plotted but too small to be visible). Lower subpanels show the orbital motion that remains after subtracting proper motion and parallax. This is for display purposes only, as our analysis jointly fits all three (proper motion, parallax, and orbital motion) simultaneously. \label{fig:orbit}}
\end{figure*}

\begin{figure*}
\centerline{
\includegraphics[width=6.5in]{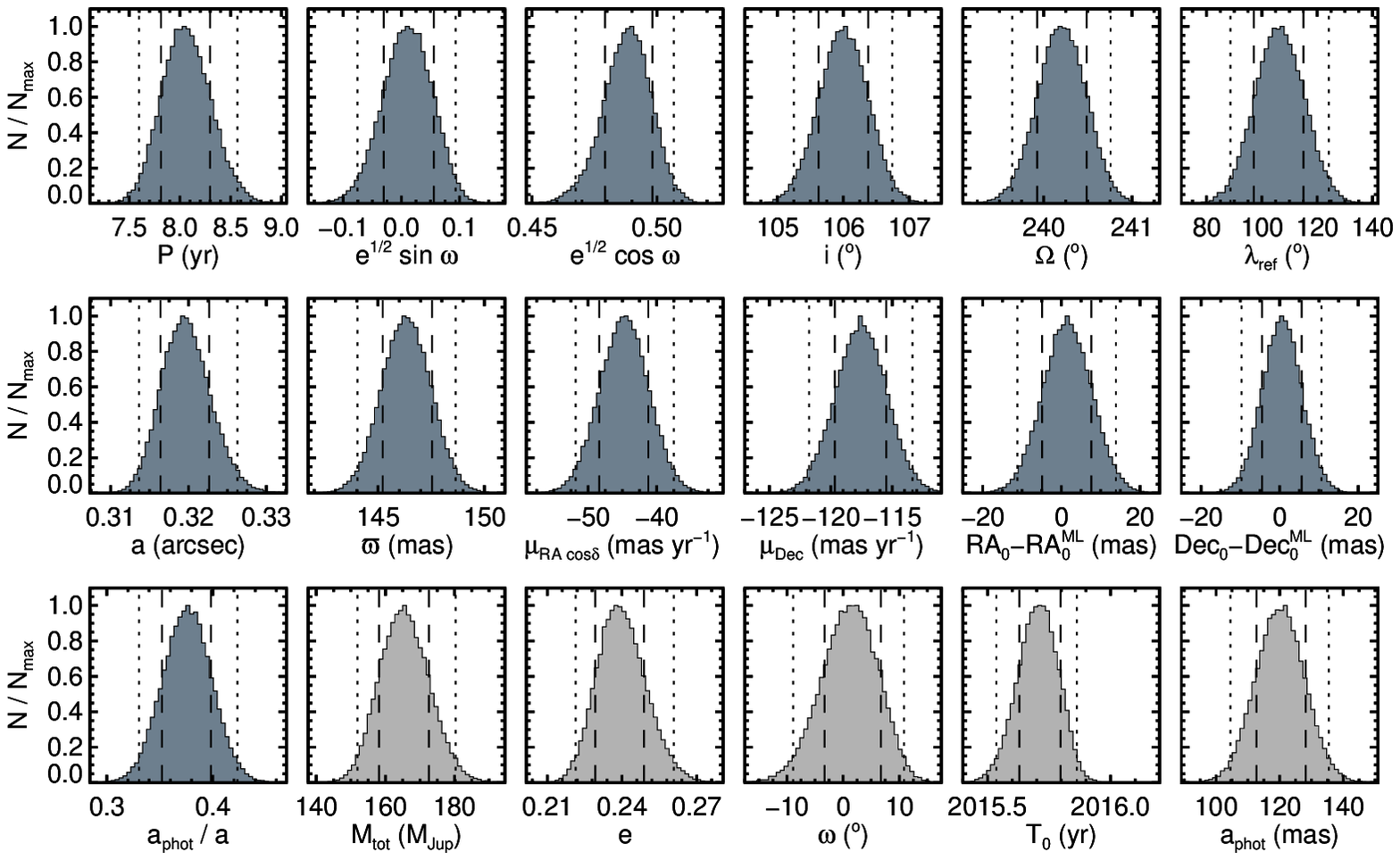}
}
\vskip -4.4 truein
\caption{Marginalized posterior distributions for all directly fitted orbital parameters in our PT-MCMC analysis (dark gray histograms). Posteriors for properties computed from the directly fitted parameters, like total mass and eccentricity, are shown in light gray histograms. \label{fig:mcmc}}
\end{figure*}

\subsection{IRTF/SpeX \label{sec:spex}}

We obtained a near-IR spectrum of \obj\ with the NASA Infrared Telescope Facility (IRTF). Our observations were taken on 2018~October~23~UT with clear skies and seeing of $\approx$0$\farcs$9. We used the facility  spectrograph SpeX \citep{2003PASP..115..362R} in the short-wavelength cross-dispersed (SXD) mode with the $0\farcs3 \times 15\arcsec$ slit ($R \approx 2000$) aligned with the parallactic angle ($327.49$\degree). We took six exposures of $120$~s each in standard ABBA pattern to achieve $> 100$ SNR per pixel in both $H$ and $K$ bands, sufficient for measuring precise metallicities of late-M dwarfs \citep[e.g.,][]{2014AJ....147..160M}. We observed the A0V standard star HD~48481 within $20$~min and 0.01 airmass of the science target for  telluric correction. We reduced the SXD spectra in a standard fashion using version 4.1 of the Spextool software package \citep{2004PASP..116..362C}.

\section{Orbit, Parallax, and Dynamical Masses \label{sec:orbit}}

We combined our Keck LGS AO relative astrometry with our integrated-light astrometry from CFHT/WIRCam in a single analysis fitting the orbit, parallax, and proper motion. We did not use the multi-bandpass averaged relative astrometry from \citet{2015AJ....150..180B} in our analysis, as that epoch (2015~Jan~11~UT) is contemporaneous with our first data (2015~Jan~14~UT). 

Our approach is very similar to our past work \citep{2015ApJ...805...56D, 2017ApJS..231...15D}. Six of the thirteen parameters are shared between the resolved and integrated-light data, all relating to the orbit: period ($P$), eccentricity ($e$) and  argument of periastron ($\omega$) parametrized as $\sqrt{e}\sin{\omega}$ and $\sqrt{e}\cos{\omega}$ for the fit in order to accommodate near-circular orbits, inclination ($i$), P.A.\ of the ascending node ($\Omega$), and mean longitude at the reference epoch ($\lambda_{\rm ref}$). The reference epoch ($t_{\rm ref}$) is defined to be 2010~January~1~00:00~UT (2455197.5~JD). There are two parameters for the size of the orbit. One is the total semimajor axis ($a$) in angular units for the resolved orbit.  The photocenter orbit size ($a_{\rm phot}$) is represented in our fit by the ratio $a_{\rm phot}/a$. The five remaining parameters are all related to the CFHT astrometry: relative parallax ($\varpi_{\rm rel}$), proper motion ($\mu_{\rm rel}$) in R.A. and decl., and the R.A. and decl.\ at the reference epoch $t_{\rm ref}$. The only parameters without uniform priors were $P$ and $a$ (log-flat), $i$ (random viewing angles, i.e., $\sin{i}$), and an approximately uniform space density prior ($\varpi_{\rm rel}^{-4}$).

We use the parallel-tempering Markov chain Monte Carlo (PT-MCMC) ensemble sampler in \texttt{emcee~v2.1.0} \citep{2013PASP..125..306F} that is based on the algorithm described by \citet{2005PCCP....7.3910E}. Our results are based on the ``coldest'' of 30 chains, where the hottest chain effectively samples all of the allowed parameter space. We use 100 walkers to sample our 13-parameter model over $8\times10^4$ steps. The initial state of the MCMC is a random, uniform draw over all of parameter space for bounded parameters ($e$, $\omega$, $\Omega$, $i$, $\lambda_{\rm ref}$; $2\,{\rm yr} < P < 2000$\,yr; $0\farcs01 < \log{a} < 1\farcs0$; $-1 < a_{\rm phot}/a < 1$); Gaussian draw of $\pm$100\,mas around the least-squares fit of the reference epoch R.A. and decl.; $\pm$30\% around the least-squares fit of the relative proper motion; and $\pm$20\% around the least-squares fit of the relative parallax. After these wide ranging initial states, the PT-MCMC converged quickly to a tightly clustered set of orbital parameters. We excluded the first 75\% of the chain as burn-in, where the last portion that was kept after we verified that it had stabilized in the mean and rms (among walkers) for each parameter.   

\begin{deluxetable*}{lccc}
\tablecaption{PT-MCMC Orbital Posteriors for WISE~J0720-0846AB \label{tbl:mcmc}}
\setlength{\tabcolsep}{0.10in}
\tabletypesize{\tiny}
\tablewidth{0pt}
\tablehead{
\colhead{Property}              &
\colhead{Median $\pm$1$\sigma$} &
\colhead{95.4\% c.i.}           &
\colhead{Prior}                 }
\startdata
\multicolumn{4}{c}{Fitted parameters} \\[1pt]
\cline{1-4}
\multicolumn{4}{c}{} \\[-5pt]
Orbital period, $P$ (yr)                                                     & $8.06_{-0.25}^{+0.24}$           &         7.60, 8.57         & $1/P$ (log-flat)                                                   \\[3pt]
Semimajor axis, $a$ (mas)                                                    & $320\pm3$                        &          314, 326          & $1/a$ (log-flat)                                                   \\[3pt]
$\sqrt{e}\sin{\omega}$                                                       & $0.01\pm0.04$                    &      $-$0.08, 0.09         & uniform                                                            \\[3pt]
$\sqrt{e}\cos{\omega}$                                                       & $0.488_{-0.009}^{+0.010}$        &        0.467, 0.507        & uniform                                                            \\[3pt]
Inclination, $i$ (\degree)                                                   & $106.0\pm0.4$                    &        105.3, 106.7        & $\sin(i)$, $0\degree < i < 180\degree$                             \\[3pt]
P.A. of the ascending node, $\Omega$ (\degree)                               & $240.21\pm0.28$                  &       239.64, 240.76       & uniform                                                            \\[3pt]
Mean longitude at $t_{\rm ref}=2455197.5$~JD, $\lambda_{\rm ref}$ (\degree)  & $106\pm9$                        &           89, 124          & uniform                                                            \\[3pt]
${\rm R.A.}_{\rm ref}-{\rm R.A.}_{\rm ref}^{\rm ML}$ (mas)                               & $1\pm6$                          &        $-$11, 14           & uniform             \\[3pt]
${\rm decl.}_{\rm ref}-{\rm decl.}_{\rm ref}^{\rm ML}$ (mas)                             & $1\pm5$                          &        $-$10, 11           & uniform            \\[3pt]
Relative proper motion in R.A., $\mu_{\rm R.A., rel}$ (\masyr)               & $-45_{-4}^{+3}$                  &        $-$52, $-$37        & uniform                                                            \\[3pt]
Relative proper motion in decl., $\mu_{\rm decl., rel}$ (\masyr)             & $-117.5_{-2.1}^{+2.0}$           &     $-$121.8, $-$113.4     & uniform                                                            \\[3pt]
Relative parallax, $\varpi_{\rm rel}$ (mas)                                  & $146.3_{-1.1}^{+1.2}$            &        144.0, 148.6        & $1/\varpi^4$                                                       \\[3pt]
Ratio of photocenter orbit to semimajor axis, $a_{\rm phot}/a$               & $0.376_{-0.024}^{+0.022}$        &        0.331, 0.423        & uniform                                                            \\[3pt]
\cline{1-4}
\multicolumn{4}{c}{} \\[-5pt]
\multicolumn{4}{c}{Computed properties} \\[1pt]
\cline{1-4}
\multicolumn{4}{c}{} \\[-5pt]
Eccentricity, $e$                                                            & $0.240_{-0.010}^{+0.009}$        &        0.221, 0.261        & \nodata                                                            \\[3pt]
Argument of periastron, $\omega$ (\degree)                                   & $1\pm5$                          &         $-$9, 11           & \nodata                                                            \\[3pt]
Time of periastron, $T_0=t_{\rm ref}-P\frac{\lambda-\omega}{360\degree}$ (JD)& $2457282_{-28}^{+33}$            &      2457219, 2457339      & \nodata                                                            \\[3pt]
Photocenter semimajor axis, $a_{\rm phot}$ (mas)                             & $120_{-7}^{+8}$                  &          105, 135          & \nodata                                                            \\[3pt]
$(a^3 P^{-2})\times10^4$ (arcsec$^3$ yr$^{-2}$)                              & $5.02\pm0.19$                    &         4.66, 5.41         & \nodata                                                            \\[3pt]
Correction to absolute R.A. proper motion, $\Delta\mu_{\rm R.A.}$ (\masyr)   & $-1.42_{-0.12}^{+0.16}$          &      $-$1.67, $-$1.06      & \nodata                                                            \\[3pt]
Correction to absolute decl. proper motion, $\Delta\mu_{\rm decl.}$ (\masyr) & $0.97_{-0.19}^{+0.20}$           &         0.58, 1.37         & \nodata                                                            \\[3pt]
Correction to absolute parallax, $\Delta\varpi$ (mas)                        & $0.77_{-0.05}^{+0.04}$           &         0.68, 0.88         & \nodata                                                            \\[3pt]
Absolute proper motion in R.A., $\mu_{\rm R.A.}$ (\masyr)                    & $-46_{-3}^{+4}$                  &        $-$53, $-$39        & \nodata                                                            \\[3pt]
Absolute proper motion in decl., $\mu_{\rm decl.}$ (\masyr)                  & $-116.5_{-2.0}^{+2.2}$           &     $-$120.8, $-$112.3     & \nodata                                                            \\[3pt]
Absolute parallax, $\varpi$ (mas)                                            & $147.1_{-1.2}^{+1.1}$            &        144.8, 149.4        & \nodata                                                            \\[3pt]
Distance, $d$ (pc)                                                           & $6.80_{-0.06}^{+0.05}$           &         6.69, 6.90         & \nodata                                                            \\[3pt]
Semimajor axis, $a$ (AU)                                                     & $2.173_{-0.029}^{+0.028}$        &        2.118, 2.230        & \nodata                                                            \\[3pt]
Total mass, $\Mtot$ (\Mjup)                                                  & $165\pm7$                        &          152, 180          & \nodata                                                           
\enddata
\tablecomments{The full 13-parameter fit has $\chi^2 = 25.8$ (21 dof), and the relative orbit has $\chi^2 = 4.84$ (7 dof). The orbit quality metrics defined by \citet{2017ApJS..231...15D} are $\delta\log\Mtot = 0.033$\,dex, $\delta{e} = 0.020$, and $\Delta{t_{\rm obs}/P} = 0.47$, indicating a high-quality orbit determination. Maximum-likelihood coordinates at the reference epoch (2010.0): ${\rm (R.A., decl.)}_{\rm ref}^{\rm ML} = (110.0135245, -08.7809247)$.}
\end{deluxetable*}

Our data is shown alongside the PT-MCMC orbit posterior in Figure~\ref{fig:orbit}. The marginalized posteriors of all of our fitted parameters, as well as some key parameters (like $e$ and $\omega$) computed from them, are shown in Figure~\ref{fig:mcmc} and summarized in Table~\ref{tbl:mcmc}. Over just $\approx$4~years of CFHT astrometric monitoring, nonlinear perturbations of up to 60\,mas are observed. This enables a precise measurement of the photocenter orbit size ($120_{-7}^{+8}$\,mas). The ratio of the photocenter orbit size to the semimajor axis ($0.376_{-0.024}^{+0.022}$) is related to the mass ratio, via the flux ratio in the CFHT bandpass. We used the $K$-band absolute magnitude of each component to compute their $K-\Kn$ color using the relations in Appendix~A.2 of \citet{2017ApJS..231...15D}, and thereby the flux ratio in the CFHT bandpass \Kn\ of $0.0246\pm0.0016$. The very small amount of flux coming from the companion makes our mass ratio quite insensitive to the exact value of the flux ratio, such as that due to the variability of the primary star. The resulting mass ratio ($M_2/M_1 = 0.67_{-0.07}^{+0.06}$) yields individual masses of $99\pm6$\,\Mjup\ and $66\pm4$\,\Mjup. Additional observational properties of the system are given in Table~\ref{tbl:prop}.

\begin{deluxetable}{lcc}
\tablecaption{Observational Properties of WISE~J0720-0846AB \label{tbl:prop}}
\tablehead{
\colhead{Property} &
\colhead{Value} &
\colhead{Ref.}}
\startdata
\multicolumn{3}{c}{Integrated light} \\[1pt]
\hline 
SpT (opt)                         & M9.5                             & B15b   \\
SpT (NIR)                         & M9.8~\fldg\                      & G15    \\
$Y$ (mag)                         &     $ 11.56\pm0.06 $             & *, B19 \\
$J$ (mag)                         &     $10.587\pm0.023$             & *, B19 \\ 
$H$ (mag)                         & \phn$ 9.982\pm0.019$             & *, B19 \\
\CHs\ (mag)                       & \phn$ 9.999\pm0.019$             & *, B19 \\
$K$ (mag)                         & \phn$ 9.446\pm0.019$             & *, B19 \\
$d$ (pc)                          & $6.80_{-0.06}^{+0.05}$           & * \\
$M$ (\Mjup)                       &        $165\pm7$                 & * \\
\fbol\ (erg\,cm$^{-2}$\,s$^{-1}$) &  $(2.28\pm0.07) \times 10^{-10}$ & * \\[1pt]
\Lbol\ (\Lsun)                    &  $(3.44\pm0.13) \times 10^{-4}$  & * \\[1pt]
\cline{1-3} \multicolumn{3}{c}{} \\[-8pt]
\multicolumn{3}{c}{\obj{A}} \\[1pt]
\hline
SpT (NIR)                         & M$9.5\pm0.5$                     & B15a   \\
$Y$ (mag)                         &     $ 11.63\pm0.06 $             & *, B19 \\
$J$ (mag)                         &     $10.684\pm0.023$             & *, B19 \\
$H$ (mag)                         &     $10.037\pm0.019$             & *, B19 \\
\CHs\ (mag)                       &     $10.066\pm0.019$             & *, B19 \\
$K$ (mag)                         & \phn$ 9.479\pm0.019$             & *, B19 \\
$M$ (\Mjup)                       &         $99\pm6$                 & * \\
\fbol\ (erg\,cm$^{-2}$\,s$^{-1}$) &  $(2.18\pm0.07) \times 10^{-10}$ & * \\[1pt]
\Lbol\ (\Lsun)                    &  $(3.29\pm0.13) \times 10^{-4}$  & * \\[1pt]
\cline{1-3} \multicolumn{3}{c}{} \\[-8pt]
\multicolumn{3}{c}{\obj{B}} \\[1pt]
\hline
SpT (IR)                          & T$5.5\pm0.5$                     & B15a   \\
$Y$ (mag)                         &     $ 14.54\pm0.07 $             & *, B19 \\
$J$ (mag)                         &     $ 13.26\pm0.04 $             & *, B19 \\
$H$ (mag)                         &     $ 13.24\pm0.04 $             & *, B19 \\
\CHs\ (mag)                       &     $ 13.05\pm0.04 $             & *, B19 \\
$K$ (mag)                         &     $ 13.31\pm0.07 $             & *, B19 \\
$M$ (\Mjup)                       &         $66\pm4$                 & * \\
\fbol\ (erg\,cm$^{-2}$\,s$^{-1}$) &  $(1.02\pm0.17) \times 10^{-11}$ & * \\[1pt]
\Lbol\ (\Lsun)                    &  $(1.5\pm0.3) \times 10^{-5}$    & * \\
\enddata
\tablerefs{(*)~this work; (B15a)~\citet{2015AJ....150..180B}; (B15b)~\citet{2015AJ....149..104B}; (B19)~Best et al.\ (2019, in prep.); (G15)~\citet{2015ApJS..219...33G}.}
\tablecomments{All photometry on the MKO system.}
\end{deluxetable}

\subsection{Comparison with Previous Work}

Our absolute parallax $147.1_{-1.2}^{+1.1}$\,mas is consistent with and considerably more precise than previously measured values of $142\pm38$\,mas from \citet{2014A&A...561A.113S}, which was based on catalog astrometry from DSS, SSS, DENIS, CMC and WISE, and $165\pm30$\,mas from \citet{2015A&A...574A..64I}, which was based on combining their data with the \citet{2014A&A...561A.113S} astrometry. \obj\ has photometry reported by \Gaia\ but no proper motion or parallax solution in DR2, possibly due to its multiplicity or very red color.

Analysis of the orbit was performed by \citet{2015AJ....150..180B} based on their two epochs of NIRC2 imaging from 2014 and 2015 as well as RV monitoring from the Lick/Hamilton and Keck/NIRSPEC spectrographs. They concluded that the orbit was quite eccentric ($e = 0.77^{+0.02}_{-0.04}$) and nearly edge on ($i = 93\fdg6^{+1.6}_{-1.4}$) with a remarkably short period ($P=4.1^{+2.7}_{-1.3}$\,yr). Our eccentricity ($e=0.234^{+0.009}_{-0.010}$) is highly inconsistent with the analysis of \citet{2015AJ....150..180B}, and our inclination and orbital period are larger by 7.5$\sigma$ and 1.4$\sigma$, respectively. The explanation for this disagreement is their 2014~Jan~19~UT epoch of astrometry, which was the original candidate detection. We did not use this data point in our orbit fit, and our MCMC posterior predicts a separation of $103\pm5$\,mas and PA of $4\degree\pm8\degree$ at that epoch. This differs from the astrometry of the candidate source identified by \citet{2015AJ....149..104B}, which was at a separation of $139\pm14$\,mas and PA of $262\degree\pm2\degree$. Our predicted separation at that epoch is much tighter and was very likely not resolvable. For comparison, we have never resolved the binary at separations tighter than $\approx$190\,mas, and at one of our unresolved observation epochs (2016~Oct~11~UT) it is predicted to have been wider than the reported 2014~Jan separation, at $108.5\pm0.6$\,mas. The FWHM of our imaging at that epoch was 51\,mas, with a Strehl ratio of 0.10, while the data from \citet{2015AJ....149..104B} had a FWHM of 230\,mas and Strehl ratio of 0.014. Therefore, we infer that the companion was not detectable in their imaging, and the candidate faint source was a PSF artifact. Visual inspection of the 2014 NIRC2 imaging obtained from the Keck Observatory Archive supports this conclusion.

\begin{figure}
\centerline{
\includegraphics[width=3.25in]{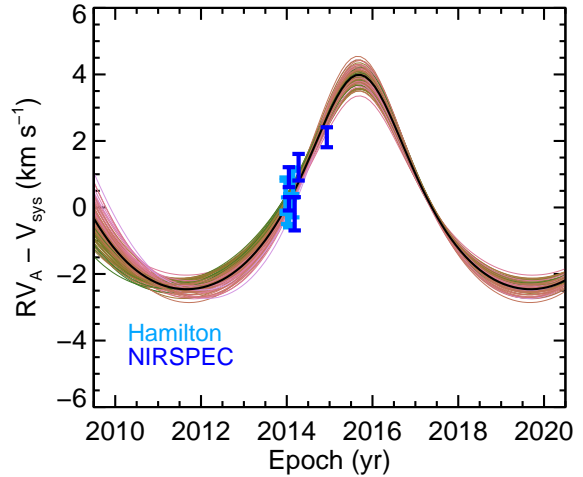}
}
\vskip 0.0 truein
\caption{The RV orbit of the late-M primary, predicted from our astrometric analysis, plotted over more than one full period alongside measurements from \citet{2015AJ....150..180B, 2015AJ....149..104B}. The mean system velocity has been subtracted off of each set of measurements. The best-fit orbit is a thick black line, and 100 orbit solutions drawn randomly from our MCMC posterior are thin lines colored according to the dynamical total mass (same as Figure~\ref{fig:orbit}). \label{fig:rv}}
\end{figure}

Our orbit predicts the RV of the primary at the measurement epochs from \citet{2015AJ....149..104B} and \citet{2015AJ....150..180B} with a precision of 0.17--0.19\,\kms, which is typically 2--3$\times$ smaller than their measurement errors. The positive slope of the RVs breaks the degeneracy between $\omega$ and $\omega+180\degree$ in the astrometric orbit (i.e., at which node the companion is going into or coming out of the plane of the sky). Assuming the preferred value for $\omega$, then for the Lick/Hamilton data the $\chi^2$ of the null hypothesis (constant RV) is 5.2 for 6 degrees of freedom (dof) and 4.2 after subtracting the orbital motion. For the re-analyzed NIRSPEC data reported by \citet{2015AJ....150..180B} (not the originally published values from \citealp{2015AJ....149..104B}), the $\chi^2$ of the null hypothesis is 35.9 (4 dof) and 10.1 after subtracting the orbital motion. This indicates that the general RV trend was detected in these data, although the RV measurement errors are likely somewhat underestimated. After subtracting the orbital motion, we compute system velocities from the two RV data sets of $82.2\pm0.4$\,\kms\ and $83.1\pm0.7$\,\kms, from which we compute a weighted average of $82.4\pm0.3$\,\kms. This is slightly smaller than the RV of $83.1\pm0.4$\,\kms\ derived by \citet{2015AJ....149..104B} because most of their measurements were obtained when the RV of the primary was $>$0\,\kms. Figure~\ref{fig:rv} displays our predicted RV orbit for the primary alongside the measurements.

\section{Spectroscopic analysis \label{sec:spec}}

We determined the bolometric flux (\fbol) of the combined light of the \obj\ system following the same procedure as \citet{2015ApJ...804...64M}, which we briefly summarize here. We simultaneously combined and absolutely calibrated the SNIFS and SpeX spectra to photometry from Gaia DR2 \citep{Evans2018}, the Two-micron All-sky Survey \citep[2MASS,][]{Skrutskie2006}, and the Wide-field Infrared Survey Explorer \citep[WISE,][]{Wright2010}, using the appropriate filter profiles and zero-points \citep{Cohen2003,dr2_filter} to generate synthetic photometry from the spectrum. We assumed zero reddening, as the target is at $\approx$7\,pc. To account for higher stellar variability in the optical, we adopted photometric errors of 0.05\,mag in the $Bp$ band and 0.04\,mag in the $G$ and $Rp$ bands. To create a full spectral energy distribution (SED), we filled in gaps in the spectral coverage (e.g., beyond 2.4\,\micron) using the best-fit BT-Settl models \citep{Allard2013}. 

\begin{figure}
    \centering
    \includegraphics[width=0.49\textwidth]{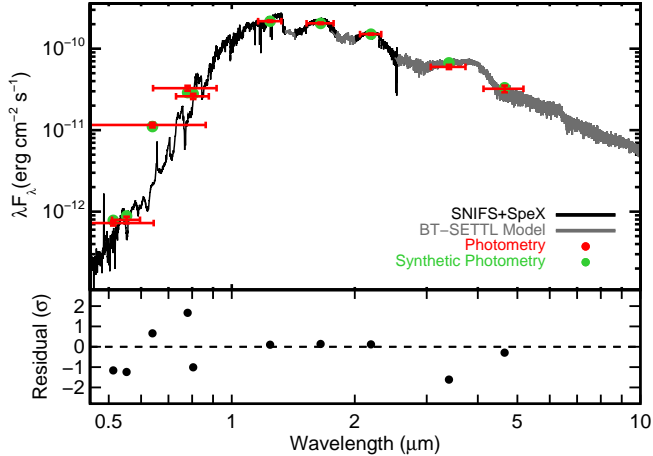}
    \caption{Top: absolutely calibrated spectrum from SNIFS and SpeX (black), best-fit BT-Settl model (gray), as well as literature (red) and synthetic (green) photometry for \obj\ in combined light. Horizontal error bars indicate the width of the filter, while vertical error bars are the measurement errors. Both the observed spectrum and photometry are expected to contain flux from the faint companion, and by fitting a model appropriate for the primary only we implicitly assume what our measured flux ratios indicate, that the companion flux is negligible in the optical and NIR. Bottom: residuals between observed and synthetic photometry in standard deviations.}
    \label{fig:sed}
\end{figure}

Figure~\ref{fig:sed} shows the observed and model spectra we used, as well as the photometric residuals. To calculate \fbol\ from the SED, we integrated the combined and calibrated spectrum over all wavelengths. Errors on \fbol\ account for measurement errors in the observed spectrum, the range of possible BT-Settl models that can fit the data, as well as errors in filter profiles and zero-points. This analysis yields an \fbol\ of $(2.28\pm0.07) \times 10^{-10}$\,erg\,cm$^{-2}$\,s$^{-1}$ for the combined light of the system.

By using a single-component model, our analysis implicitly assumes that the contribution of the T~dwarf companion to the combined-light SED is negligible relative to photometric and BT-Settl model uncertainties. This is certainly the case in the optical, and conversely it is unlikely to be true at long wavelengths, which is why we did not include $W3$ and $W4$ in our analysis. 

In the NIR, we must consider the binary separation and the fact that we used a $0\farcs3$ slit for our observations. Using our orbit determination, we predict the separation and PA of the binary at this epoch was $348.7\pm1.6$\,mas and $67.69\pm0.11$\degree. This is nearly orthogonal to the parallactic angle-aligned slit used for our observations (100.20\degree\ different). Assuming that the slit was centered on the primary, and a Gaussian PSF with a FWHM of $0\farcs9$, the companion contribution to our spectrum was 32\% smaller than if the slit were aligned with the binary PA. Thus, the companion contribution is still quite significant, so we performed tests fitting two component SEDs to our data, one for each binary component instead of just one for the primary, and calculated the impact on our final derived \fbol. The BT-Settl models we used for these tests do not reproduce the companion photometry well, possibly due to deficiencies in the atmospheric physics, and the model grid is only coarsely sampled in \Teff, making accurate SED fitting unfeasible. Therefore, while the two-component fits are not suitable for accurate SED modeling, they do provide a estimate of the error introduced by assuming a single-component model for the SED fitting. We find that including a wide range of models still resulted in at most $\pm$1\% of variation in the derived \fbol. This is negligible compared to our \fbol\ measurement error of 3.1\%.

\begin{figure}
    \centerline{
    \includegraphics[width=0.45\textwidth]{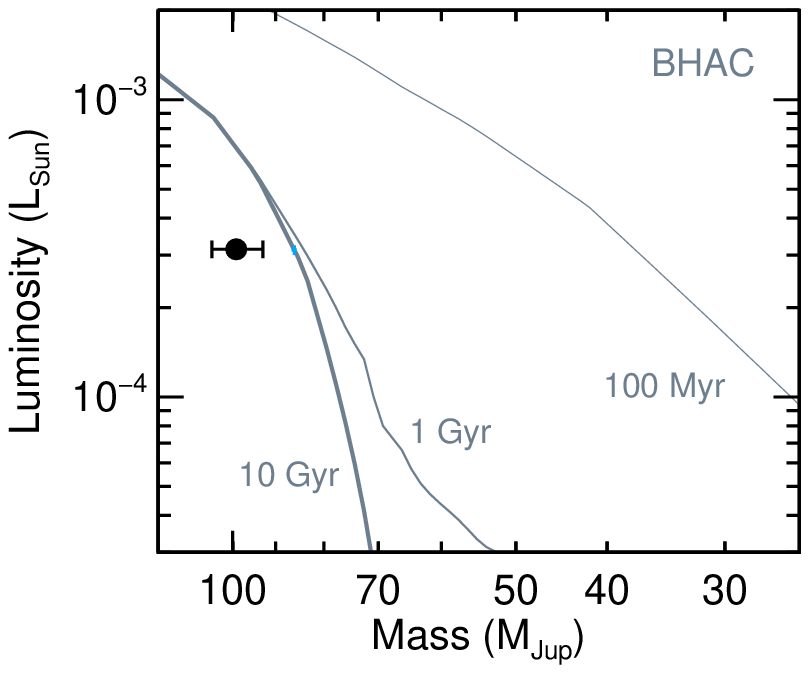}
    }
    \centerline{
    \includegraphics[width=0.45\textwidth]{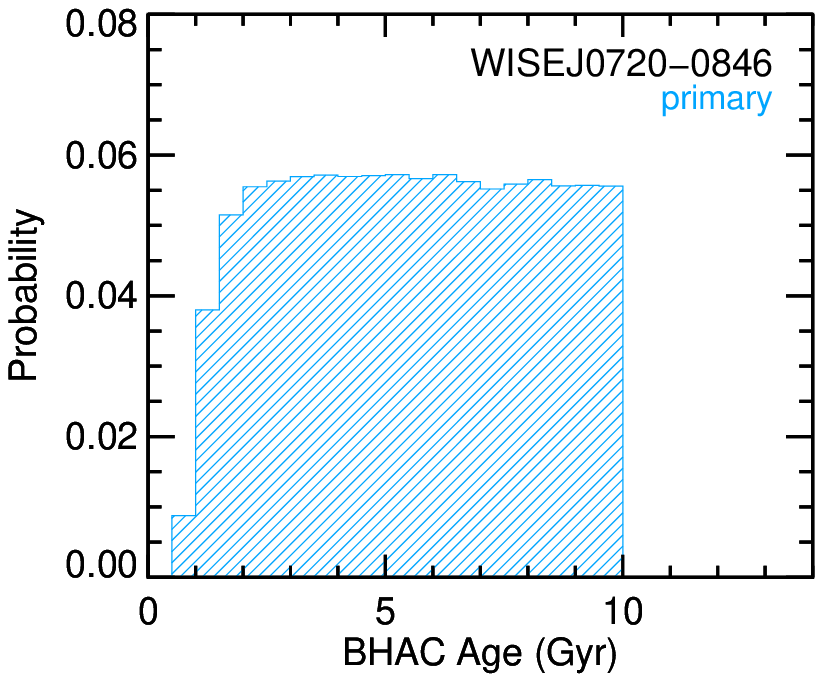}
    }
    \caption{Top: Our measured mass and luminosity for \obj{A} shown alongside the BHAC15 isochrones that are appropriate for a late-M dwarf. \obj{A} appears massive for its luminosity and is marginally inconsistent with models ($2.1\sigma$). Bottom: Age distribution from our individual-mass analysis of \obj{A}. Given that we determine \obj{A} to be a star, its age is consistent with the full range of main-sequence ages covered by the BHAC15 models ($<10$\,Gyr), and the shape of the distribution is simply the result of our uniform age prior.}
    \label{fig:evol-pri}
\end{figure}

Our combined-light \fbol\ and measured parallax yield a total bolometric luminosity of $(3.44\pm0.13)\times10^{-4}$\,\Lsun. We computed the bolometric luminosity of the T~dwarf companion from the $K$-band absolute magnitude relation of \citet{2017ApJS..231...15D}, finding $\Lbol = (1.5\pm0.3)\times10^{-5}$\,\Lsun, which is 4.4\% of the total bolometric flux. Converting this back to $\fbol = (1.02\pm0.17)\times10^{-11}$\,erg\,cm$^{-2}$\,s$^{-1}$, and subtracting it from the combined-light flux, gives $\fbol = (2.18\pm0.07)\times10^{-10}$\,erg\,cm$^{-2}$\,s$^{-1}$ and $\Lbol = (3.29\pm0.13)\times10^{-4}$\,\Lsun\ for the primary component.

From our SpeX spectrum, we also derived a metallicity of ${\rm [Fe/H]} = +0.15\pm0.10$\,dex using the calibration of \citet{Mann2014}. It is based on the correlation between metallicity and the strength of atomic Na, Ca, and K lines in NIR spectra of M dwarfs, validated using wide binaries containing a FGK primary and a late-M companion \citep[e.g.,][]{Bonfils:2005,Rojas-Ayala:2012uq}. Given the lack of these same alkali features in mid-T dwarf spectra, and the faintness of the T~dwarf companion here, the influence of the companion on our metallicity analysis is expected to be negligible.

\section{Evolutionary Model Analysis} \label{sec:evol}

In order to derive additional fundamental parameters for the system components, we have combined our measured masses and luminosities with evolutionary models. We perform a rejection sampling analysis that is the same as we used in \citet{2017ApJS..231...15D}. Briefly, we start with randomly drawn masses and luminosities from our measured distributions, tracking the covariance in these parameters, and combine these with randomly drawn ages according to a uniform prior in time. Each random draw corresponds to a measured luminosity as well as a model-derived luminosity (from mass and age), and the rejection probability is computed from the difference between these two luminosities. Over three iterations we adjust the range over which ages are drawn as needed to ensure a well-sampled posterior on model-derived properties. In addition to this ``individual-mass'' analysis, we also perform a ``total-mass'' analysis where only the resolved luminosities and dynamical total mass are input as measurements. In this case random component masses and a system age are drawn for each trial, and the rejection probability is computed from the component luminosities and summed mass.

\begin{figure}
    \centerline{
    \includegraphics[width=0.45\textwidth]{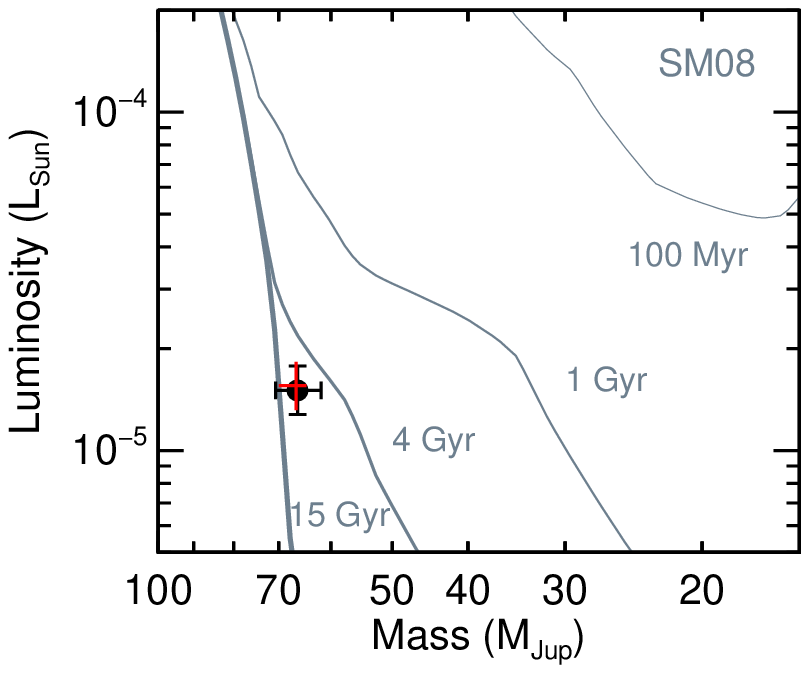}
    }
    \centerline{
    \includegraphics[width=0.45\textwidth]{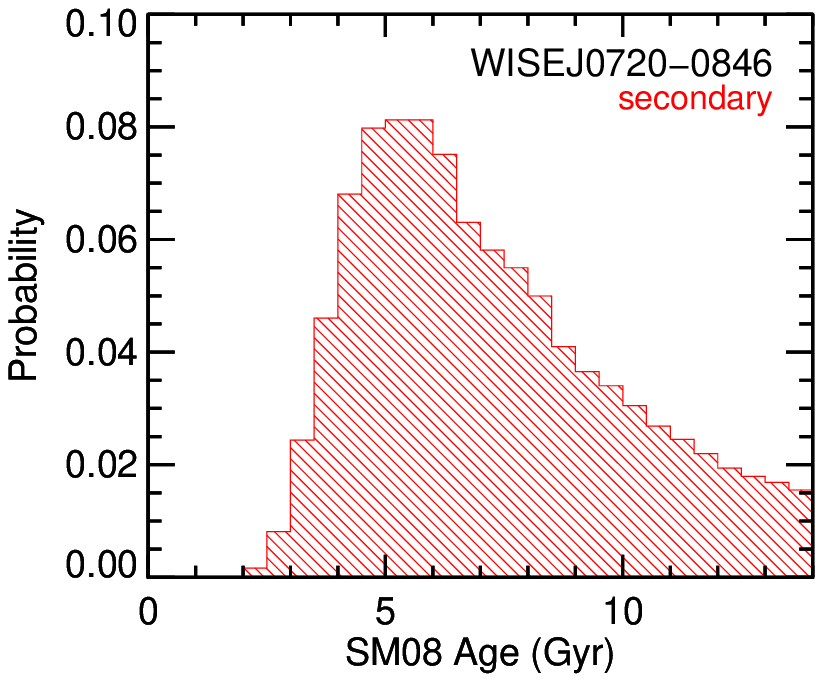}
    }
    \caption{Top: Our measured mass and luminosity for \obj{B} shown alongside the SM08 hybrid isochrones, which are appropriate for a mid-T dwarf. \obj{B} is rather massive for its luminosity but consistent with models at old ages. The colored symbol indicates the 1$\sigma$ posterior on mass and luminosity after our rejection sampling analysis. Bottom: Age distribution from our individual-mass analysis of \obj{B}, which is based on our input mass and luminosity measurements and a uniform prior on age. Given how broad the distribution is, it is strongly influenced by our prior.}
    \label{fig:evol-sec}
\end{figure}

\begin{deluxetable*}{lcccccccc}
\tablecaption{Fundamental Properties of WISE~J0720-0846AB \label{tbl:evol}}
\tablehead{
\colhead{}                            &
\colhead{}                            &
\multicolumn{3}{c}{Using Total Mass}  &
\colhead{}                            &
\multicolumn{3}{c}{Using Individual Masses}  \\ \cline{3-5} \cline{7-9}
\colhead{Property}                    &
\colhead{}                            &
\colhead{Primary}                     &
\colhead{Secondary}                   &
\colhead{$\Delta = {\rm B} - {\rm A}$}&
\colhead{}                            &
\colhead{Primary}                     &
\colhead{Secondary}                   &
\colhead{$\Delta = {\rm B} - {\rm A}$}}
\startdata
\multicolumn{9}{c}{Input Observed Properties} \\
\cline{1-9}
Mass $M$ (\Mjup)        & & \multicolumn{2}{c}{$165\pm7$                          }                   & \nodata                             & & $99\pm6$                            & $66\pm4$                            & $-33\pm8$                           \\
Mass ratio $q$          & & \nodata                             & \nodata                             & \nodata                             & & \multicolumn{2}{c}{$0.67_{-0.07}^{+0.06}$             }                   & \nodata                             \\
log(\Lbol) [\Lsun]      & & $-3.503\pm0.016$                    & $-4.82\pm0.07$                      & $-1.32\pm0.08$                      & & \nodata                             & \nodata                             & \nodata                             \\
\cline{1-9}
\multicolumn{9}{c}{}\\[-6pt] 
\multicolumn{9}{c}{Derived from BHAC15 Evolutionary Models} \\ 
\cline{1-9} 
Mass $M$ (\Mjup)        & & \nodata & \nodata           & \nodata & &  $86.0_{-0.5}^{+0.6}$                & \nodata                     & \nodata \\
log(\Lbol) [\Lsun]      & & \nodata & \nodata           & \nodata & &  $-3.505_{-0.017}^{+0.016}$          & \nodata                     & \nodata \\
Age $t$ (Gyr)           & & \multicolumn{2}{c}{\nodata} & \nodata & &  $5.5_{-3.2}^{+2.9}$                 & \nodata                     & \nodata \\
$\log(t)$ [yr]          & & \multicolumn{2}{c}{\nodata} & \nodata & &  $9.74_{-0.15}^{+0.26}$              & \nodata                     & \nodata \\
\Teff\ (K)              & & \nodata & \nodata           & \nodata & &  $2407_{-15}^{+14}$                  & \nodata                     & \nodata \\
Radius (\Rjup)          & & \nodata & \nodata           & \nodata & &  $0.992_{-0.007}^{+0.006}$           & \nodata                     & \nodata \\
\logg\ [cm\,s$^{-2}$]   & & \nodata & \nodata           & \nodata & &  $5.3363_{-0.0029}^{+0.0036}$        & \nodata                     & \nodata \\
log(Li/Li$_{\rm init}$) & & \nodata & \nodata           & \nodata & &  $<-4.0$                             & \nodata                     & \nodata \\
MKO($J-K$) (mag)        & & \nodata & \nodata           & \nodata & &  $0.735\pm0.006$                     & \nodata                     & \nodata \\
MKO($J-H$) (mag)        & & \nodata & \nodata           & \nodata & &  $0.448\pm0.005$                     & \nodata                     & \nodata \\
\cline{1-9} 
\multicolumn{9}{c}{}\\[-6pt] 
\multicolumn{9}{c}{Derived from SM08 Hybrid Evolutionary Models} \\ 
\cline{1-9} 
Mass $M$ (\Mjup)        & & \nodata & \nodata           & \nodata & & \nodata                             & $66.5_{-2.0}^{+3.5}$                & \nodata \\
log(\Lbol) [\Lsun]      & & \nodata & \nodata           & \nodata & & \nodata                             & $-4.81\pm0.07$                      & \nodata \\
Age $t$ (Gyr)           & & \multicolumn{2}{c}{\nodata} & \nodata & & \nodata                             & $6.8_{-3.1}^{+2.2}$                 & \nodata \\
$\log(t)$ [yr]          & & \multicolumn{2}{c}{\nodata} & \nodata & & \nodata                             & $9.83_{-0.19}^{+0.17}$              & \nodata \\
\Teff\ (K)              & & \nodata & \nodata           & \nodata & & \nodata                             & $1250\pm40$                         & \nodata \\
Radius (\Rjup)          & & \nodata & \nodata           & \nodata & & \nodata                             & $0.822_{-0.016}^{+0.015}$           & \nodata \\
\logg\ [cm\,s$^{-2}$]   & & \nodata & \nodata           & \nodata & & \nodata                             & $5.387_{-0.023}^{+0.033}$           & \nodata \\
MKO($J-K$) (mag)        & & \nodata & \nodata           & \nodata & & \nodata                             & $0.802_{-0.406}^{+0.366}$           & \nodata \\
MKO($J-H$) (mag)        & & \nodata & \nodata           & \nodata & & \nodata                             & $0.638_{-0.180}^{+0.176}$           & \nodata \\
\cline{1-9} 
\multicolumn{9}{c}{}\\[-6pt] 
\multicolumn{9}{c}{Derived from Cond Evolutionary Models} \\ 
\cline{1-9} 
Mass $M$ (\Mjup)        & & $86.7\pm0.5$                        & $70.7_{-1.7}^{+2.4}$                & $-16.1_{-1.8}^{+2.6}$               & & $86.7_{-0.5}^{+0.6}$                & $67.6_{-2.9}^{+4.2}$                & $-19_{-3}^{+4}$                     \\
log(\Lbol) [\Lsun]      & & $-3.506_{-0.016}^{+0.017}$          & $-4.82\pm0.07$                      & $-1.31\pm0.07$                      & & $-3.506_{-0.017}^{+0.015}$          & $-4.83\pm0.07$                      & $-1.32\pm0.07$                      \\
Mass ratio $q$          & & \multicolumn{2}{c}{$0.815_{-0.020}^{+0.029}$          }                   & \nodata                             & & \multicolumn{2}{c}{$0.78_{-0.03}^{+0.05}$             }                   & \nodata                             \\
Age $t$ (Gyr)           & & \multicolumn{2}{c}{$7.4_{-1.1}^{+2.5}$                }                   & \nodata                             & & $5.4_{-3.5}^{+2.6}$                 & $5.3_{-2.5}^{+1.5}$                 & $0\pm3$                             \\
$\log(t)$ [yr]          & & \multicolumn{2}{c}{$9.87_{-0.06}^{+0.13}$             }                   & \nodata                             & & $9.74_{-0.15}^{+0.26}$              & $9.73_{-0.15}^{+0.19}$              & $0.01_{-0.32}^{+0.27}$              \\
\Teff\ (K)              & & $2404\pm15$                         & $1290\pm50$                         & $-1110\pm50$                        & & $2404_{-15}^{+14}$                  & $1270\pm50$                         & $-1130\pm50$                        \\
Radius (\Rjup)          & & $1.000\pm0.006$                     & $0.767_{-0.008}^{+0.013}$           & $-0.232_{-0.012}^{+0.011}$          & & $1.000\pm0.006$                     & $0.776_{-0.014}^{+0.013}$           & $-0.223_{-0.017}^{+0.014}$          \\
\logg\ [cm\,s$^{-2}$]   & & $5.333\pm0.003$                     & $5.473_{-0.010}^{+0.023}$           & $0.140_{-0.013}^{+0.024}$           & & $5.3323_{-0.0026}^{+0.0038}$        & $5.447_{-0.021}^{+0.049}$           & $0.115_{-0.027}^{+0.045}$           \\
log(Li/Li$_{\rm init}$) & & \nodata                             & \nodata                             & \nodata                             & & \nodata                             & $-0.0210\pm0.0013$                  & \nodata                             \\
MKO($J-K$) (mag)        & & $0.649\pm0.005$                     & $-0.166_{-0.044}^{+0.052}$          & $-0.815_{-0.045}^{+0.052}$          & & $0.650_{-0.006}^{+0.004}$           & $-0.167_{-0.053}^{+0.056}$          & $-0.816_{-0.053}^{+0.057}$          \\
MKO($J-H$) (mag)        & & $0.316\pm0.004$                     & $-0.283_{-0.023}^{+0.019}$          & $-0.599_{-0.022}^{+0.021}$          & & $0.317\pm0.004$                     & $-0.287_{-0.024}^{+0.023}$          & $-0.603\pm0.024$                    \\
\enddata 
\tablecomments{The BHAC models do not extend to the luminosity of the secondary, so for the BHAC individual-mass analysis only the results for \obj{A} are given.  The SM08 models do not extend to the luminosity of the primary, so for the SM08 individual-mass analysis only the results for \obj{B} are given.}
\end{deluxetable*} 

Table~\ref{tbl:evol} shows the results of our evolutionary model analysis, which we performed for three different sets of models. The \citet[][hereinafter BHAC15]{2015A&A...577A..42B} grid encompasses the luminosity of \obj{A} but not its much cooler companion, so we only report individual-mass analysis for the primary with these models. Conversely, the \citet[][hereinafter SM08]{2008ApJ...689.1327S} models cover the cooler \obj{B} but not the primary. We use the ``hybrid'' version of the SM08 models, that transition from cloudy to cloud-free atmospheres as objects cool from 1400\,K to 1200\,K, as they are more appropriate for an object like \obj{B} that is on the blue end of the L/T transition and may still possess clouds. Finally, we include the ``Cond'' models \citep{2003A&A...402..701B} because they are the only models to cover the physical properties of both components However, unlike the above models, Cond does not include condensate clouds in the photosphere and thus should not be physically appropriate for either component.

\obj{A} has a high mass for its luminosity, making it marginally inconsistent with models (2.1$\sigma$, Figure~\ref{fig:evol-pri}). Future refinement of the total mass and, more importantly, mass ratio could resolve this discrepancy. For example, our total-mass analysis that does not use the mass ratio gives a self-consistent result with a slightly lower mass for the primary and higher mass for the secondary, making the mass ratio 0.81, which is exactly at the edge of our 95.4\% credible interval of 0.55--0.81.

\begin{deluxetable*}{lccccccccc}
\tablecaption{Dynamical Mass Measurements for T~Dwarfs \label{tbl:t-mass}}
\tablehead{
\colhead{Name}    &
\colhead{Spectral}    &
\colhead{Mass}    &
\colhead{Semimajor}    &
\colhead{Eccentricity}    &
\colhead{Mass ratio}    &
\colhead{Distance}        &
\colhead{$\log(\Lbol/\Lsun)$}        &
\colhead{\Teff}        &
\colhead{Ref.}     \\[-10pt]
\colhead{}         &
\colhead{Type}         &
\colhead{(\Mjup)}  &
\colhead{axis (au)} &
\colhead{}      &
\colhead{$M_2/M_1$}      &
\colhead{(pc)}         &
\colhead{(dex)}         &
\colhead{(K)}         &
\colhead{}       }
\startdata
\object{HD 4747B}                                      & T$1\pm2$     & $66.2^{+2.5}_{-3.0}$    & $10.1^{+0.4}_{-0.5}$      & $0.735\pm0.003$              & $0.077^{+0.004}_{-0.005}$ & $18.79\pm0.04$          & $ -4.55\pm0.08 $ & $1380\pm50$ & BJ18, Cr16, Cr18, * \\
\object{$\epsilon$ Indi B}                             & T$1\pm1$     & $68.0\pm0.9$            & \nodata                   & \nodata                      & \nodata                   & $3.6386\pm0.0033$       & $-4.699\pm0.017$ & $1312\pm9\phn$  & Ki10, Ca12, BJ18, * \\
\object{**LUH 16B}                                     & T$0.5\pm1.0$ & $28.55^{+0.26}_{-0.25}$ & $3.557^{+0.026}_{-0.023}$ & $0.343\pm0.005$              & $0.8519\pm0.0024$         & $1.994\pm0.0003$        & $ -4.71\pm0.10 $ & $1190\pm60$ & Bu13, Ga17, LS18, * \\
\object[SDSSJ105213.51+442255.7]{SDSS J1052+4422B}    & T$1.5\pm1.0$ & $39.4^{+2.6}_{-2.7}$    & $1.86\pm0.03$             & $0.1399_{-0.0023}^{+0.0022}$ & $0.78\pm0.07$             & $26.2\pm0.4$            & $ -4.64\pm0.07 $ & $1270\pm40$ & Du15, DL17 \\
\object[SDSSpJ042348.57-041403.5]{SDSS J0423$-$0414B}   & T$2.0\pm0.5$ & $31.8^{+1.5}_{-1.6}$    & $2.291^{+0.0027}_{-0.0028}$ & $0.272_{-0.007}^{+0.008}$    & $0.62\pm0.04$             & $14.07_{-0.17}^{+0.16}$ & $ -4.72\pm0.07 $ & $1200\pm40$ & DL17 \\
\object[DENISJ225210.7-173013]{DENIS J2252$-$1730B}    & T$3.5\pm0.5$ & $41\pm4$                & $1.95\pm0.04$             & $0.334\pm0.009$              & $0.70^{+0.08}_{-0.09}$    & $15.9\pm0.3$            & $ -4.76\pm0.07 $ & $1210\pm50$ & DL17 \\
\object[2MASSJ15344984-2952274A]{2MASS J1534$-$2952A}  & T$4.5\pm0.5$ & $51\pm5$                & \nodata                   & \nodata                      & \nodata                   & $15.9\pm0.3$            & $ -4.91\pm0.07 $ & $1150\pm50$ & Li08, DL17 \\
\object[2MASSJ14044948-3159330]{2MASS J1404$-$3159B}   & T$5.0\pm0.5$ & $55^{+6}_{-7}$          & $3.15^{+0.09}_{-0.11}$    & $0.825\pm0.005$              & $0.84\pm0.06$             & $23.5\pm0.6$            & $ -4.87\pm0.07 $ & $1190\pm50$ & DL17 \\
\obj{B}                                                & T$5.5\pm0.5$ & $66\pm4$                & $2.17\pm0.03$             & $0.240_{-0.010}^{+0.009}$    & $0.67^{+0.06}_{-0.07}$    & $6.80_{-0.06}^{+0.05}$  & $ -4.82\pm0.07 $ & $1250\pm40$ & Bu15, * \\
\object[2MASSJ15344984-2952274B]{2MASS J1534$-$2952B}  & T$5.0\pm0.5$ & $48\pm5$                & $3.40\pm0.06$             & $0.0027^{+0.0028}_{-0.0027}$ & $0.95^{+0.13}_{-0.16}$    & $15.9\pm0.3$            & $ -4.99\pm0.07 $ & $1100\pm50$ & Li08, DL17 \\
\object{$\epsilon$ Indi C}                             & T$6\pm1$     & $53.1\pm0.3$            & $2.4214\pm0.0013$         & $0.5401\pm0.0007$            & $0.781\pm0.014$           & $3.6386\pm0.0033$       & $-5.232\pm0.020$ & $ 975\pm11$ & Ki10, Ca12, BJ18 \\
\object{Gl 758B}                                       & T8?          & $38.1^{+1.7}_{-1.5}$    & $30^{+5}_{-8}$            & $0.40\pm0.09$                & $0.048^{+0.011}_{-0.015}$ & $15.603\pm0.005$        & $ -6.07\pm0.03 $ & $ 594\pm10$ & Vi16, Bo18, BJ18, Br19 \\
\object{HD 4113C}                                      & T9?          & $66^{+5}_{-4}$          & $23^{+4}_{-3}$            & $0.38^{+0.08}_{-0.06}$       & $0.060^{+0.005}_{-0.004}$ & $41.87\pm0.09$          & $  -6.0\pm0.1  $ & $ 700\pm40$ & BJ18, Ch18, *
\enddata
\tablerefs{(*)---this work; 
BJ18---\citet{2018AJ....156...58B}; 
Bo18---\citet{2018AJ....155..159B}; 
Br18---\citet{2018arXiv181107285B}; 
Bu13---\citet{2013ApJ...772..129B};
Bu15---\citet{2015AJ....150..180B}
Ca12---\citet{2012CardosoC};  
Cr16---\citet{2016ApJ...831..136C};
Cr18---\citet{2018ApJ...853..192C};
Ch18---\citet{2018AA...614A..16C};
Du15---\citet{2015ApJ...805...56D}; 
DL17---\citet{2017ApJS..231...15D}; 
Ga17---\citet{2017ApJ...846...97G}; 
Ki10---\citet{2010AA...510A..99K}; 
Li08---\citet{2008ApJ...689..436L};
LS18---\citet{2018AA...618A.111L};
Vi16---\citet{2016AA...587A..55V}.}
\tablecomments{For objects not in DL17, we report \Teff\ computed using the SM08 hybrid models and the same method as in DL17. Only HD~4113C's mass depends on an assumption for its host-star mass, and we estimate its \Lbol\ to be the same as \object{UGPS~J072227.51$-$054031.2} \citep{2015ApJ...810..158F}, but with larger uncertainty, given their identical $J$-band absolute magnitudes.}
\end{deluxetable*}

Given that \obj{A}'s mass and luminosity  is above the theoretical substellar boundary, our model analysis yields no age information from it. \obj{B} on the other hand is only consistent with relatively old ages ($\gtrsim$4\,Gyr), although even this quantitative age limit is influenced by our prior assumption of uniform age given the broad output posterior (Figure~\ref{fig:evol-sec}). 

Table~\ref{tbl:evol} gives physical properties derived from our evolutionary model analysis, including our posteriors on the masses and luminosities. Every Monte Carlo trial preserved in our rejection sampling analysis corresponds to a part of parameter space actually covered by models, so our posterior primary mass of $86.0_{-0.5}^{+0.6}$\,\Mjup\ (BHAC15) is significantly smaller, with smaller errors, than our input measurement. The primary's model-derived effective temperature (\Teff) of $2407_{-15}^{+14}$\,K is consistent with its spectral type of M9.5, as the mass-calibrated spectral type--\Teff\ relation of \citet{2017ApJS..231...15D} based on the same models gives $2400\pm90$\,K. Likewise, the SM08-based relation gives $\Teff = 1180\pm80$\,K for \obj{B}, which agrees with our mass-calibrated value of $1250\pm40$\,K. The smaller radius of the secondary ($0.822_{-0.016}^{+0.015}$\,\Rjup) relative to the primary ($0.992_{-0.007}^{+0.006}$\,\Rjup) results in the secondary having a slightly higher surface gravity ($\logg = 5.387_{-0.023}^{+0.033}$\,dex) than the primary ($\logg = 5.336_{-0.003}^{+0.004}$\,dex), despite its lower mass.


\section{Discussion} \label{sec:discussion}

\subsection{Substellar Boundary \label{sec:hbmm}}

The minimum stellar mass for sustained fusion of hydrogen is of interest to understanding stellar interiors, and it is key in determining which objects are viable hosts of habitable planets. This threshold is not a sharp function of mass, as progressively lower mass stars take correspondingly longer times to reach the main sequence. For example, in the BHAC15 models a 75-\Mjup\ object is clearly a star but takes $\approx$2\,Gyr to reach its main-sequence luminosity of $\approx8\times10^{-5}$\,\Lsun. Evolutionary models differ in their detailed predictions of this mass boundary, ranging from $\approx$70\,\Mjup\ (the highest mass SM08 object that is clearly substellar) on the low end, 73--75\,\Mjup\ \citep{2000ARA&A..38..337C}, 73--79\,\Mjup\ \citep{2001RvMP...73..719B, 2011ApJ...736...47B}, and as high as 82--83\,\Mjup\ \citep{2019ApJ...879...94F}. Models also predict a modest dependence on metallicity for the mass boundary. For instance, according to \citet{2019ApJ...879...94F} there is a $\approx$0.04\,dex decrease in the mass boundary per dex of increasing metallicity. Thus, low-metallicity brown dwarfs could be more massive than typical field objects, such as T~subdwarfs \citep[e.g.,][]{2014MNRAS.440..359B,2019MNRAS.486.1260Z}.

One property that is consistently predicted across all models that objects less luminous than $\approx3\times10^{-5}$\,\Lsun\ at an age of $\approx$10\,Gyr are on a path to fade and cool forever, and some models predict an even higher luminosity for this boundary. The luminosity of \obj{B} is $(1.5\pm0.3)\times10^{-5}$\,\Lsun, well below this conservative, theoretical threshold for a substellar object. Its mass of $66\pm4$\,\Mjup\ is therefore rather high, but not inconsistent with even the lowest model-predicted mass boundary.

There are relatively few empirical determinations of the mass of the substellar boundary. \citet{2017ApJS..231...15D} used the first large sample of objects with individual dynamical masses that have luminosities ranging from $\sim$10$^{-3}$--$10^{-5}$\,\Lsun\ to determine the substellar boundary as the mass for which objects diverged from a one-to-one relation between mass and \Lbol. They found a boundary of $\approx$70\,\Mjup\ based on a lack of low-\Lbol\ objects at higher masses than that.  The uncertainty on this determination is still to be established, but preliminary work from \citet{2018AAS...23134934C} suggests that it is approximately $\pm$4\,\Mjup. 

\begin{figure}
    \centerline{\includegraphics[width=0.45\textwidth]{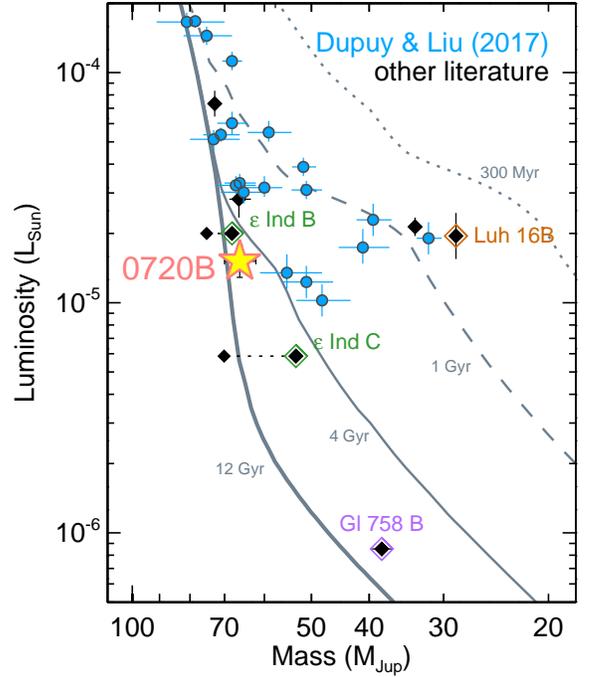}}
    \caption{Luminosity as a function of mass for ultracool dwarfs that have model-independent mass measurements, with gray lines showing SM08 hybrid evolutionary model isochrones. Most measurements come from \citet[][blue circles]{2017ApJS..231...15D}, and other literature measurements are plotted as black diamonds. Notable literature T~dwarfs are highlighted with colored diamonds: $\epsilon$~Ind~B and C (green), \object[**LUH16B]{Luhman~16B} (brown), and Gl~758B (purple). For $\epsilon$~Ind~BC, we plot both the lower masses from \citet{2012CardosoC} and higher masses from \citet{2018ApJ...865...28D}, connected by horizontal dotted lines. Like $\epsilon$~Ind~B, our mass measurement for \obj{B} is higher than other objects of comparable luminosity.}
    \label{fig:m-l}
\end{figure}

Individual, very cool objects can place a lower limit on the substellar boundary. The T1.5 primary in $\epsilon$~Ind~BC has a luminosity of $(2.00\pm0.08)\times10^{-5}$\,\Lsun\ \citep{2010A&A...510A..99K} and two distinct mass measurements in the literature. \citet{2012CardosoC} obtained a mass of $68.0\pm0.9$\,\Mjup\ based on relative astrometry from VLT/NACO and absolute astrometry from VLT/FORS2. In constrast, \citet{2018ApJ...865...28D} found a mass of $75.0\pm0.8$\,\Mjup\ by combining their photocenter (unresolved) orbit measured in  optical imaging from CTIOPI \citep{2005AJ....129.1954J} and du~Pont/CAPScam \citep{2009PASP..121.1218B} with a portion of the VLT/NACO relative astrometry also used by \citet{2012CardosoC}. Both masses are consistent with the boundary at $70\pm4$~\Mjup\ from \citet{2017ApJS..231...15D}, even though there is strong tension in the mass results for $\epsilon$~Ind~B. We note that \citet{2018ApJ...865...28D} report a total mass of $144.5\pm1.1$\,\Mjup\ that relies on both unresolved and resolved astrometry (but independent of the binary's flux ratio), while \citet{2012CardosoC} report $121.2\pm1.1$\,\Mjup\ based on the relative astrometry alone. Almost all orbit parameters  have significant tension between the two results, but the only ones directly relevant to the total mass are semimajor axis (8.3\% higher in \citealp{2018ApJ...865...28D}) and period (3.6\% higher in \citealp{2018ApJ...865...28D}). Given the dependence of $\Mtot \propto a^3 P^{-2}$, this explains the 19\% difference in total mass, and it is mostly driven by the difference in semimajor axis. \citet{2012CardosoC} measured $665.5\pm0.6$\,mas compared to $721\pm8$\,mas from \citet{2018ApJ...865...28D}. It is not obvious what could cause such a large difference in semimajor axis (55\,mas, $\approx$4 NACO pixels) between these two works based on the same relative astrometry data. Systematic errors of 8\% in relative astrometry would defy explanation, especially as the astrometry of \citet{2012CardosoC} was calibrated using images of a wide binary yielding typical precision of 2--3\,mas. Likewise, 8\% systematics in the absolute astrometric orbit ($\sim$15\,mas) seem unlikely, though perhaps somewhat more plausible as the per-epoch precision from \citet{2018ApJ...865...28D} was 2--6\,mas. Their analysis did not report a goodness-of-fit metric, so we cannot assess whether their errors were appropriate. There is no good explanation for this discrepancy at the moment, but $\epsilon$~Ind~B generally lends credence to the existence of ``massive'' T~dwarfs in the near-solar metallicity field population, consistent with our results for \obj{B}.

\subsection{Solar System Encounter \label{sec:encounter}}

We have used our new proper motion, parallax, and orbit-corrected system velocity of $82.4\pm0.3$\,\kms\ to compute an updated space motion for \obj{AB}. We find $(U,V,W) = (-57.48\pm0.22, -59.18\pm0.21, 0.22\pm0.13)$\,\kms\ and $(X,Y,Z) = (-4.89\pm0.04, -4.72\pm0.04, 0.271\pm0.002)$\,pc. \citet{2015ApJ...800L..17M} showed that over timescales of $\sim$100\,kyr, a linear trajectory is the same within 2.5\% accuracy as compared to more detailed calculations that include the effects of the Galactic potential. Using a linear trajectory, we find that the system's closest passage to the Sun was $0.333\pm0.010$\,pc ($68.7\pm2.0$\,kAU) at an epoch $80.5\pm0.7$\,kya. Our results are in good agreement with the analysis of \citet{2015ApJ...800L..17M}, with our errors being about an order of magnitude smaller thanks to our improved parallax and proper motion measurements. Our results conclusively rule out the hypothesis that \obj{AB} could have passed through the inner Oort cloud ($<$20\,kAU; \citealp{1981AJ.....86.1730H}) but would have instead passed through the outer Oort cloud where comets can have stable orbits ($\lesssim$100\,kAU; e.g., \citealp{1984Natur.311...38S}).

\subsection{Comparison to $M$--$M_K$--[Fe/H] Relation \label{sec:mass-mag}}

\citet{2019ApJ...871...63M} have produced the most precise mass--magnitude relation to date for K and M dwarfs, which can be used to derive masses accurate to $\approx$3\% above the substellar boundary. We computed the mass posterior using their code\footnote{\url{https://github.com/awmann/M_-M_K-}} 
and an input 2MASS \Ks-band apparent magnitude of $9.504\pm0.020$\,mag derived from the $K-\Ks$ color--absolute magnitude relations from Appendix~A.1 of \citet{2017ApJS..231...15D}. The \citet{2019ApJ...871...63M} relation gives a mass of $85.0\pm2.2$\,\Mjup\ from just its magnitude and distance, $84.9\pm2.3$\,\Mjup\ if we also provide its metallicity of [Fe/H]~=~$0.15\pm0.10$\,dex (Section~\ref{sec:spec}).

Our measured mass for \obj{A} is $99\pm6$\,\Mjup, which is marginally inconsistent with (2.1$\sigma$ higher than) the mass derived from the empirical relation. Given that \obj{B} also appears to be somewhat massive for a mid-T dwarf, this could be a hint that the true total mass is closer to the low end of our posterior distribution. It is also possible that the true mass ratio could be closer to the high end of our posterior distribution, which would shift more of the mass out of the primary and into the secondary, so the secondary would still be relatively massive for its spectral type. \citet{2017ApJS..231...15D} found a mean mass of 36\,\Mjup\ for five T2--T5.5 dwarfs (rms 9\,\Mjup), which is expectedly lower than we find for \obj{B} given the younger mean age of the field population sample used by \citet{2017ApJS..231...15D}. Currently, the dominant source of uncertainty in the total mass is the orbital period ($8.06_{-0.25}^{+0.24}$\,yr). Given how short this period is, its error should drop rapidly in the next few years, thereby reducing the mass uncertainties and potentially resolving the marginal discrepancy in the primary mass. The mass ratio depends strongly on the photocenter orbit size, and this parameter is still quite covariant with the proper motion of the system, and a longer time baseline will also greatly help in reducing this degeneracy.

\section{Summary}

We present new astrometry and relative photometry from Keck and CFHT, including $K$- and \Lp-band imaging from Keck's new pyramid wavefront sensor. We jointly fit our observations with a 13-parameter orbit and parallax solution that yields precise individual dynamical masses and a greatly improved distance measurement. (Like many binaries, \obj\ does not have a \Gaia~DR2 parallax.) The mass of the primary is marginally inconsistent with the empirical mass--magnitude relation and stellar models. The brown dwarf companion is rather massive ($66\pm4$\,\Mjup) compared to other mid-T dwarfs with dynamical masses, which may be partly explained by the current uncertainty in the orbital period and thereby total mass. This mass is consistent with evolutionary models, within the errors, assuming the system is several Gyr old. Such an age is consistent with past interpretation of the stellar host properties and space motion. Finally, our much more precise parallax and proper motion, along with our first accurate orbit determination, enable a more rigorous assessment of the system's recent close encounter with the solar system.

\acknowledgments

T.J.D.\ acknowledges research support from Gemini Observatory, which is operated by the Association of Universities for Research in Astronomy, Inc., on behalf of the international Gemini partnership of Argentina, Brazil, Canada, Chile, the Republic of Korea, and the United States of America.
M.C.L.\ acknowledges National Science Foundation (NSF) grant AST-1518339.
The near-infrared pyramid wavefront sensor is supported by NSF grant AST-1611623.  
The camera used with the pyramid wavefront sensor is provided by Don Hall with support from the National Science Foundation under grant AST-1106391.
M.A.T. acknowledges support from the DOE CSGF through grant DE-SC0019323. 
Our research has employed NASA ADS; SIMBAD; VizieR; and J.~R.~A.\ Davenport's IDL implementation of the cubehelix color scheme \citep{2011BASI...39..289G}.

\facility{Keck:II (LGS AO, NIRC2, PyWFS AO), CFHT (WIRCam), IRTF (SpeX), UH:2.2m (SNIFS)}

\software{{\tt Spextool} \citep{2004PASP..116..362C}, {\tt emcee} \citep{2013PASP..125..306F}, {\tt M\_-M\_K-} \citep{2019ApJ...871...63M}}

\bibliographystyle{aasjournal}
\bibliography{refs.bib}

\begin{thebibliography}{}
\expandafter\ifx\csname natexlab\endcsname\relax\def\natexlab#1{#1}\fi
\providecommand{\url}[1]{\href{#1}{#1}}
\providecommand{\dodoi}[1]{doi:~\href{http://doi.org/#1}{\nolinkurl{#1}}}
\providecommand{\doeprint}[1]{\href{http://ascl.net/#1}{\nolinkurl{http://ascl.net/#1}}}
\providecommand{\doarXiv}[1]{\href{https://arxiv.org/abs/#1}{\nolinkurl{https://arxiv.org/abs/#1}}}

\bibitem[{{Aldering} {et~al.}(2002){Aldering}, {Adam}, {Antilogus}, {Astier},
  {Bacon}, {Bongard}, {Bonnaud}, {Copin}, {Hardin}, {Henault}, {Howell},
  {Lemonnier}, {Levy}, {Loken}, {Nugent}, {Pain}, {Pecontal}, {Pecontal},
  {Perlmutter}, {Quimby}, {Schahmaneche}, {Smadja}, \&
  {Wood-Vasey}}]{Aldering2002}
{Aldering}, G., {Adam}, G., {Antilogus}, P., {et~al.} 2002, in Society of
  Photo-Optical Instrumentation Engineers (SPIE) Conference Series, Vol. 4836,
  Survey and Other Telescope Technologies and Discoveries, ed. J.~A. {Tyson} \&
  S.~{Wolff}, 61--72

\bibitem[{{Allard} {et~al.}(2013){Allard}, {Homeier}, {Freytag},
  {Schaffenberger}, {}, \& {Rajpurohit}}]{Allard2013}
{Allard}, F., {Homeier}, D., {Freytag}, B., {et~al.} 2013, Memorie della
  Societa Astronomica Italiana Supplementi, 24, 128

\bibitem[{{Bacon} {et~al.}(2001){Bacon}, {Copin}, {Monnet}, {Miller},
  {Allington-Smith}, {Bureau}, {Carollo}, {Davies}, {Emsellem}, {Kuntschner},
  {Peletier}, {Verolme}, \& {de Zeeuw}}]{Bacon2001}
{Bacon}, R., {Copin}, Y., {Monnet}, G., {et~al.} 2001, \mnras, 326, 23

\bibitem[{{Bailer-Jones}(2018)}]{2018A&A...609A...8B}
{Bailer-Jones}, C.~A.~L. 2018, \aap, 609, A8,
  \dodoi{10.1051/0004-6361/201731453}

\bibitem[{{Bailer-Jones} {et~al.}(2018){Bailer-Jones}, {Rybizki}, {Fouesneau},
  {Mantelet}, \& {Andrae}}]{2018AJ....156...58B}
{Bailer-Jones}, C.~A.~L., {Rybizki}, J., {Fouesneau}, M., {Mantelet}, G., \&
  {Andrae}, R. 2018, \aj, 156, 58, \dodoi{10.3847/1538-3881/aacb21}

\bibitem[{Baraffe {et~al.}(2003)Baraffe, {Chabrier}, {Barman}, {Allard}, \&
  {Hauschildt}}]{2003A&A...402..701B}
Baraffe, I., {Chabrier}, G., {Barman}, T.~S., {Allard}, F., \& {Hauschildt},
  P.~H. 2003, \aap, 402, 701, \dodoi{10.1051/0004-6361:20030252}

\bibitem[{Baraffe {et~al.}(2015)Baraffe, {Homeier}, {Allard}, \&
  {Chabrier}}]{2015A&A...577A..42B}
Baraffe, I., {Homeier}, D., {Allard}, F., \& {Chabrier}, G. 2015, \aap, 577,
  A42, \dodoi{10.1051/0004-6361/201425481}

\bibitem[{{Bardalez Gagliuffi} {et~al.}(2014){Bardalez Gagliuffi}, {Burgasser},
  {Gelino}, {Looper}, {Nicholls}, {Schmidt}, {Cruz}, {West}, {Gizis}, \&
  {Metchev}}]{2014ApJ...794..143B}
{Bardalez Gagliuffi}, D.~C., {Burgasser}, A.~J., {Gelino}, C.~R., {et~al.}
  2014, \apj, 794, 143, \dodoi{10.1088/0004-637X/794/2/143}

\bibitem[{Bertin \& {Arnouts}(1996)}]{1996A&AS..117..393B}
Bertin, E., \& {Arnouts}, S. 1996, \aaps, 117, 393

\bibitem[{{Bond} {et~al.}(2018){Bond}, {Wizinowich}, {Chun}, {Mawet}, {Lilley},
  {Cetre}, {Jovanovic}, {Delorme}, {Wetherell}, {Jacobson}, {Lockhart},
  {Warmbier}, {Wallace}, {Hall}, {Goebel}, {Guyon}, {Plantet}, {Agapito},
  {Giordano}, {Esposito}, \& {Femenia-Castella}}]{2018SPIE10703E..1ZB}
{Bond}, C.~Z., {Wizinowich}, P., {Chun}, M., {et~al.} 2018, in Society of
  Photo-Optical Instrumentation Engineers (SPIE) Conference Series, Vol. 10703,
  Adaptive Optics Systems VI, 107031Z

\bibitem[{{Bonfils} {et~al.}(2005){Bonfils}, {Delfosse}, {Udry}, {Santos},
  {Forveille}, \& {S{\'e}gransan}}]{Bonfils:2005}
{Bonfils}, X., {Delfosse}, X., {Udry}, S., {et~al.} 2005, \aap, 442, 635

\bibitem[{{Boss} {et~al.}(2009){Boss}, {Weinberger}, {Anglada-Escud{\'e}},
  {Thompson}, {Burley}, {Birk}, {Pravdo}, {Shaklan}, {Gatewood}, {Majewski}, \&
  {Patterson}}]{2009PASP..121.1218B}
{Boss}, A.~P., {Weinberger}, A.~J., {Anglada-Escud{\'e}}, G., {et~al.} 2009,
  \pasp, 121, 1218, \dodoi{10.1086/647960}

\bibitem[{Bouchez {et~al.}(2004)Bouchez, {Le Mignant}, {van Dam}, {Chin},
  {Hartman}, {Johansson}, {Lafon}, {Stomski}, {Summers}, \&
  {Wizinowich}}]{2004SPIE.5490..321B}
Bouchez, A.~H., {Le Mignant}, D., {van Dam}, M.~A., {et~al.} 2004, in Society
  of Photo-Optical Instrumentation Engineers (SPIE) Conference Series, Vol.
  5490, Advancements in Adaptive Optics, ed. D.~{Bonaccini Calia}, B.~L.
  {Ellerbroek}, \& R.~{Ragazzoni}, 321--330

\bibitem[{{Bowler} {et~al.}(2018){Bowler}, {Dupuy}, {Endl}, {Cochran},
  {MacQueen}, {Fulton}, {Petigura}, {Howard}, {Hirsch}, {Kratter}, {Crepp},
  {Biller}, {Johnson}, \& {Wittenmyer}}]{2018AJ....155..159B}
{Bowler}, B.~P., {Dupuy}, T.~J., {Endl}, M., {et~al.} 2018, \aj, 155, 159,
  \dodoi{10.3847/1538-3881/aab2a6}

\bibitem[{{Brandt} {et~al.}(2018){Brandt}, {Dupuy}, \&
  {Bowler}}]{2018arXiv181107285B}
{Brandt}, T.~D., {Dupuy}, T.~J., \& {Bowler}, B.~P. 2018, arXiv e-prints,
  arXiv:1811.07285.
\newblock \doarXiv{1811.07285}

\bibitem[{{Burgasser} {et~al.}(2015{\natexlab{a}}){Burgasser}, {Melis}, {Todd},
  {Gelino}, {Hallinan}, \& {Bardalez Gagliuffi}}]{2015AJ....150..180B}
{Burgasser}, A.~J., {Melis}, C., {Todd}, J., {et~al.} 2015{\natexlab{a}}, \aj,
  150, 180, \dodoi{10.1088/0004-6256/150/6/180}

\bibitem[{{Burgasser} {et~al.}(2013){Burgasser}, {Sheppard}, \&
  {Luhman}}]{2013ApJ...772..129B}
{Burgasser}, A.~J., {Sheppard}, S.~S., \& {Luhman}, K.~L. 2013, \apj, 772, 129,
  \dodoi{10.1088/0004-637X/772/2/129}

\bibitem[{{Burgasser} {et~al.}(2015{\natexlab{b}}){Burgasser}, {Gillon},
  {Melis}, {Bowler}, {Michelsen}, {Bardalez Gagliuffi}, {Gelino}, {Jehin},
  {Delrez}, {Manfroid}, \& {Blake}}]{2015AJ....149..104B}
{Burgasser}, A.~J., {Gillon}, M., {Melis}, C., {et~al.} 2015{\natexlab{b}},
  \aj, 149, 104, \dodoi{10.1088/0004-6256/149/3/104}

\bibitem[{{Burningham} {et~al.}(2014){Burningham}, {Smith}, {Cardoso}, {Lucas},
  {Burgasser}, {Jones}, \& {Smart}}]{2014MNRAS.440..359B}
{Burningham}, B., {Smith}, L., {Cardoso}, C.~V., {et~al.} 2014, \mnras, 440,
  359, \dodoi{10.1093/mnras/stu184}

\bibitem[{{Burrows} {et~al.}(2011){Burrows}, {Heng}, \&
  {Nampaisarn}}]{2011ApJ...736...47B}
{Burrows}, A., {Heng}, K., \& {Nampaisarn}, T. 2011, \apj, 736, 47,
  \dodoi{10.1088/0004-637X/736/1/47}

\bibitem[{Burrows {et~al.}(2001)Burrows, {Hubbard}, {Lunine}, \&
  {Liebert}}]{2001RvMP...73..719B}
Burrows, A., {Hubbard}, W.~B., {Lunine}, J.~I., \& {Liebert}, J. 2001, Reviews
  of Modern Physics, 73, 719, \dodoi{10.1103/RevModPhys.73.719}

\bibitem[{{Buton} {et~al.}(2013){Buton}, {Copin}, {Aldering}, {Antilogus},
  {Aragon}, {Bailey}, {Baltay}, {Bongard}, {Canto}, {Cellier-Holzem},
  {Childress}, {Chotard}, {Fakhouri}, {Gangler}, {Guy}, {Hsiao}, {Kerschhaggl},
  {Kowalski}, {Loken}, {Nugent}, {Paech}, {Pain}, {P{\'e}contal}, {Pereira},
  {Perlmutter}, {Rabinowitz}, {Rigault}, {Runge}, {Scalzo}, {Smadja}, {Tao},
  {Thomas}, {Weaver}, {Wu}, \& {Nearby SuperNova Factory}}]{Buton2013}
{Buton}, C., {Copin}, Y., {Aldering}, G., {et~al.} 2013, \aap, 549, A8

\bibitem[{{Cancino} \& {Dupuy}(2018)}]{2018AAS...23134934C}
{Cancino}, A.~A., \& {Dupuy}, T. 2018, in American Astronomical Society Meeting
  Abstracts, Vol. 231, American Astronomical Society Meeting Abstracts \#231,
  349.34

\bibitem[{Cardoso(2012)}]{2012CardosoC}
Cardoso, C. V.~V. 2012, PhD thesis, University of Exeter

\bibitem[{Chabrier \& {Baraffe}(2000)}]{2000ARA&A..38..337C}
Chabrier, G., \& {Baraffe}, I. 2000, \araa, 38, 337,
  \dodoi{10.1146/annurev.astro.38.1.337}

\bibitem[{{Cheetham} {et~al.}(2018){Cheetham}, {S{\'e}gransan}, {Peretti},
  {Delisle}, {Hagelberg}, {Beuzit}, {Forveille}, {Marmier}, {Udry}, \&
  {Wildi}}]{2018AA...614A..16C}
{Cheetham}, A., {S{\'e}gransan}, D., {Peretti}, S., {et~al.} 2018, \aap, 614,
  A16, \dodoi{10.1051/0004-6361/201630136}

\bibitem[{{Cohen} {et~al.}(2003){Cohen}, {Wheaton}, \& {Megeath}}]{Cohen2003}
{Cohen}, M., {Wheaton}, W.~A., \& {Megeath}, S.~T. 2003, \aj, 126, 1090

\bibitem[{{Crepp} {et~al.}(2016){Crepp}, {Gonzales}, {Bechter}, {Montet},
  {Johnson}, {Piskorz}, {Howard}, \& {Isaacson}}]{2016ApJ...831..136C}
{Crepp}, J.~R., {Gonzales}, E.~J., {Bechter}, E.~B., {et~al.} 2016, \apj, 831,
  136, \dodoi{10.3847/0004-637X/831/2/136}

\bibitem[{{Crepp} {et~al.}(2018){Crepp}, {Principe}, {Wolff}, {Giorla Godfrey},
  {Rice}, {Cieza}, {Pueyo}, {Bechter}, \& {Gonzales}}]{2018ApJ...853..192C}
{Crepp}, J.~R., {Principe}, D.~A., {Wolff}, S., {et~al.} 2018, \apj, 853, 192,
  \dodoi{10.3847/1538-4357/aaa2fd}

\bibitem[{Cushing {et~al.}(2004)Cushing, {Vacca}, \&
  {Rayner}}]{2004PASP..116..362C}
Cushing, M.~C., {Vacca}, W.~D., \& {Rayner}, J.~T. 2004, \pasp, 116, 362,
  \dodoi{10.1086/382907}

\bibitem[{Cutri {et~al.}(2003)Cutri, {Skrutskie}, {van Dyk}, {Beichman},
  {Carpenter}, {Chester}, {Cambresy}, {Evans}, {Fowler}, {Gizis}, {Howard},
  {Huchra}, {Jarrett}, {Kopan}, {Kirkpatrick}, {Light}, {Marsh}, {McCallon},
  {Schneider}, {Stiening}, {Sykes}, {Weinberg}, {Wheaton}, {Wheelock}, \&
  {Zacarias}}]{2003tmc..book.....C}
Cutri, R.~M., {Skrutskie}, M.~F., {van Dyk}, S., {et~al.} 2003, {2MASS All Sky
  Catalog of point sources.} (The IRSA 2MASS All-Sky Point Source Catalog,
  NASA/IPAC Infrared Science
  Archive.~http://irsa.ipac.caltech.edu/applications/Gator/)

\bibitem[{{de la Fuente Marcos} \& {de la Fuente
  Marcos}(2018)}]{2018RNAAS...2b..30D}
{de la Fuente Marcos}, R., \& {de la Fuente Marcos}, C. 2018, Research Notes of
  the American Astronomical Society, 2, 30, \dodoi{10.3847/2515-5172/aac2d0}

\bibitem[{{Dieterich} {et~al.}(2018){Dieterich}, {Weinberger}, {Boss}, {Henry},
  {Jao}, {Gagn{\'e}}, {Astraatmadja}, {Thompson}, \&
  {Anglada-Escud{\'e}}}]{2018ApJ...865...28D}
{Dieterich}, S.~B., {Weinberger}, A.~J., {Boss}, A.~P., {et~al.} 2018, \apj,
  865, 28, \dodoi{10.3847/1538-4357/aadadc}

\bibitem[{Diolaiti {et~al.}(2000)Diolaiti, {Bendinelli}, {Bonaccini}, {Close},
  {Currie}, \& {Parmeggiani}}]{2000A&AS..147..335D}
Diolaiti, E., {Bendinelli}, O., {Bonaccini}, D., {et~al.} 2000, \aaps, 147, 335

\bibitem[{Dupuy {et~al.}(2016)Dupuy, {Kratter}, {Kraus}, {Isaacson}, {Mann},
  {Ireland}, {Howard}, \& {Huber}}]{2016ApJ...817...80D}
Dupuy, T.~J., {Kratter}, K.~M., {Kraus}, A.~L., {et~al.} 2016, \apj, 817, 80,
  \dodoi{10.3847/0004-637X/817/1/80}

\bibitem[{Dupuy \& {Liu}(2012)}]{2012ApJS..201...19D}
Dupuy, T.~J., \& {Liu}, M.~C. 2012, \apjs, 201, 19,
  \dodoi{10.1088/0067-0049/201/2/19}

\bibitem[{{Dupuy} \& {Liu}(2017)}]{2017ApJS..231...15D}
{Dupuy}, T.~J., \& {Liu}, M.~C. 2017, \apjs, 231, 15,
  \dodoi{10.3847/1538-4365/aa5e4c}

\bibitem[{Dupuy {et~al.}(2010)Dupuy, {Liu}, {Bowler}, {Cushing}, {Helling},
  {Witte}, \& {Hauschildt}}]{2010ApJ...721.1725D}
Dupuy, T.~J., {Liu}, M.~C., {Bowler}, B.~P., {et~al.} 2010, \apj, 721, 1725,
  \dodoi{10.1088/0004-637X/721/2/1725}

\bibitem[{Dupuy {et~al.}(2015)Dupuy, {Liu}, {Leggett}, {Ireland}, {Chiu}, \&
  {Golimowski}}]{2015ApJ...805...56D}
Dupuy, T.~J., {Liu}, M.~C., {Leggett}, S.~K., {et~al.} 2015, \apj, 805, 56,
  \dodoi{10.1088/0004-637X/805/1/56}

\bibitem[{{Earl} \& {Deem}(2005)}]{2005PCCP....7.3910E}
{Earl}, D.~J., \& {Deem}, M.~W. 2005, Physical Chemistry Chemical Physics
  (Incorporating Faraday Transactions), 7, 3910, \dodoi{10.1039/B509983H}

\bibitem[{{Evans} {et~al.}(2018){Evans}, {Riello}, {De Angeli}, {Carrasco},
  {Montegriffo}, {Fabricius}, {Jordi}, {Palaversa}, {Diener}, {Busso},
  {Cacciari}, {van Leeuwen}, {Burgess}, {Davidson}, {Harrison}, {Hodgkin},
  {Pancino}, {Richards}, {Altavilla}, {Balaguer-N{\'u}{\~n}ez}, {Barstow},
  {Bellazzini}, {Brown}, {Castellani}, {Cocozza}, {De Luise}, {Delgado},
  {Ducourant}, {Galleti}, {Gilmore}, {Giuffrida}, {Holl}, {Kewley}, {Koposov},
  {Marinoni}, {Marrese}, {Osborne}, {Piersimoni}, {Portell}, {Pulone},
  {Ragaini}, {Sanna}, {Terrett}, {Walton}, {Wevers}, \&
  {Wyrzykowski}}]{Evans2018}
{Evans}, D.~W., {Riello}, M., {De Angeli}, F., {et~al.} 2018, \aap, 616, A4

\bibitem[{{Fernandes} {et~al.}(2019){Fernandes}, {Van Grootel}, {Salmon},
  {Aringer}, {Burgasser}, {Scuflaire}, {Brassard}, \&
  {Fontaine}}]{2019ApJ...879...94F}
{Fernandes}, C.~S., {Van Grootel}, V., {Salmon}, S. J.~A.~J., {et~al.} 2019,
  \apj, 879, 94, \dodoi{10.3847/1538-4357/ab2333}

\bibitem[{Filippazzo {et~al.}(2015)Filippazzo, {Rice}, {Faherty}, {Cruz}, {Van
  Gordon}, \& {Looper}}]{2015ApJ...810..158F}
Filippazzo, J.~C., {Rice}, E.~L., {Faherty}, J., {et~al.} 2015, \apj, 810, 158,
  \dodoi{10.1088/0004-637X/810/2/158}

\bibitem[{Foreman-Mackey {et~al.}(2013)Foreman-Mackey, {Hogg}, {Lang}, \&
  {Goodman}}]{2013PASP..125..306F}
Foreman-Mackey, D., {Hogg}, D.~W., {Lang}, D., \& {Goodman}, J. 2013, \pasp,
  125, 306, \dodoi{10.1086/670067}

\bibitem[{{Gagn{\'e}} {et~al.}(2015){Gagn{\'e}}, {Faherty}, {Cruz},
  {Lafreni{\'e}re}, {Doyon}, {Malo}, {Burgasser}, {Naud}, {Artigau},
  {Bouchard}, {Gizis}, \& {Albert}}]{2015ApJS..219...33G}
{Gagn{\'e}}, J., {Faherty}, J.~K., {Cruz}, K.~L., {et~al.} 2015, \apjs, 219,
  33, \dodoi{10.1088/0067-0049/219/2/33}

\bibitem[{{Gaidos} {et~al.}(2014){Gaidos}, {Mann}, {L{\'e}pine}, {Buccino},
  {James}, {Ansdell}, {Petrucci}, {Mauas}, \& {Hilton}}]{Gaidos2014}
{Gaidos}, E., {Mann}, A.~W., {L{\'e}pine}, S., {et~al.} 2014, \mnras, 443, 2561

\bibitem[{{Garcia} {et~al.}(2017){Garcia}, {Ammons}, {Salama}, {Crossfield},
  {Bendek}, {Chilcote}, {Garrel}, {Graham}, {Kalas}, {Konopacky}, {Lu},
  {Macintosh}, {Marin}, {Marois}, {Nielsen}, {Neichel}, {Pham}, {De Rosa},
  {Ryan}, {Service}, \& {Sivo}}]{2017ApJ...846...97G}
{Garcia}, E.~V., {Ammons}, S.~M., {Salama}, M., {et~al.} 2017, \apj, 846, 97,
  \dodoi{10.3847/1538-4357/aa844f}

\bibitem[{{Green}(2011)}]{2011BASI...39..289G}
{Green}, D.~A. 2011, Bulletin of the Astronomical Society of India, 39, 289.
\newblock \doarXiv{1108.5083}

\bibitem[{{Hills}(1981)}]{1981AJ.....86.1730H}
{Hills}, J.~G. 1981, \aj, 86, 1730, \dodoi{10.1086/113058}

\bibitem[{Ivanov {et~al.}(2015)Ivanov, {Vaisanen}, {Kniazev}, {Beletsky},
  {Mamajek}, {Mu{\v{z}}i{\'c}}, {Beam{\'\i}n}, {Boffin}, {Pourbaix}, {Gand hi},
  {Gulbis}, {Monaco}, {Saviane}, {Kurtev}, {Mawet}, {Borissova}, \&
  {Minniti}}]{2015A&A...574A..64I}
Ivanov, V.~D., {Vaisanen}, P., {Kniazev}, A.~Y., {et~al.} 2015, \aap, 574, A64,
  \dodoi{10.1051/0004-6361/201424883}

\bibitem[{{Jao} {et~al.}(2005){Jao}, {Henry}, {Subasavage}, {Brown}, {Ianna},
  {Bartlett}, {Costa}, \& {M{\'e}ndez}}]{2005AJ....129.1954J}
{Jao}, W.-C., {Henry}, T.~J., {Subasavage}, J.~P., {et~al.} 2005, \aj, 129,
  1954, \dodoi{10.1086/428489}

\bibitem[{{King} {et~al.}(2010){King}, {McCaughrean}, {Homeier}, {Allard},
  {Scholz}, \& {Lodieu}}]{2010AA...510A..99K}
{King}, R.~R., {McCaughrean}, M.~J., {Homeier}, D., {et~al.} 2010, \aap, 510,
  A99, \dodoi{10.1051/0004-6361/200912981}

\bibitem[{King {et~al.}(2010)King, {McCaughrean}, {Homeier}, {Allard},
  {Scholz}, \& {Lodieu}}]{2010A&A...510A..99K}
King, R.~R., {McCaughrean}, M.~J., {Homeier}, D., {et~al.} 2010, \aap, 510,
  A99, \dodoi{10.1051/0004-6361/200912981}

\bibitem[{{Lantz} {et~al.}(2004){Lantz}, {Aldering}, {Antilogus}, {Bonnaud},
  {Capoani}, {Castera}, {Copin}, {Dubet}, {Gangler}, {Henault}, {Lemonnier},
  {Pain}, {Pecontal}, {Pecontal}, \& {Smadja}}]{Lantz2004}
{Lantz}, B., {Aldering}, G., {Antilogus}, P., {et~al.} 2004, in Society of
  Photo-Optical Instrumentation Engineers (SPIE) Conference Series, Vol. 5249,
  Optical Design and Engineering, ed. L.~{Mazuray}, P.~J. {Rogers}, \&
  R.~{Wartmann}, 146--155

\bibitem[{{Lazorenko} \& {Sahlmann}(2018)}]{2018AA...618A.111L}
{Lazorenko}, P.~F., \& {Sahlmann}, J. 2018, \aap, 618, A111,
  \dodoi{10.1051/0004-6361/201833626}

\bibitem[{{Liu} {et~al.}(2016){Liu}, {Dupuy}, \&
  {Allers}}]{2016ApJ...833...96L}
{Liu}, M.~C., {Dupuy}, T.~J., \& {Allers}, K.~N. 2016, \apj, 833, 96,
  \dodoi{10.3847/1538-4357/833/1/96}

\bibitem[{Liu {et~al.}(2008)Liu, {Dupuy}, \& {Ireland}}]{2008ApJ...689..436L}
Liu, M.~C., {Dupuy}, T.~J., \& {Ireland}, M.~J. 2008, \apj, 689, 436,
  \dodoi{10.1086/591837}

\bibitem[{{Liu} {et~al.}(2010){Liu}, {Dupuy}, \&
  {Leggett}}]{2010ApJ...722..311L}
{Liu}, M.~C., {Dupuy}, T.~J., \& {Leggett}, S.~K. 2010, \apj, 722, 311,
  \dodoi{10.1088/0004-637X/722/1/311}

\bibitem[{{Ma{\'\i}z Apell{\'a}niz} \& {Weiler}(2018)}]{dr2_filter}
{Ma{\'\i}z Apell{\'a}niz}, J., \& {Weiler}, M. 2018, \aap, 619, A180

\bibitem[{{Mamajek} {et~al.}(2015){Mamajek}, {Barenfeld}, {Ivanov}, {Kniazev},
  {V{\"a}is{\"a}nen}, {Beletsky}, \& {Boffin}}]{2015ApJ...800L..17M}
{Mamajek}, E.~E., {Barenfeld}, S.~A., {Ivanov}, V.~D., {et~al.} 2015, \apjl,
  800, L17, \dodoi{10.1088/2041-8205/800/1/L17}

\bibitem[{Mann {et~al.}(2014)Mann, {Deacon}, {Gaidos}, {Ansdell}, {Brewer},
  {Liu}, {Magnier}, \& {Aller}}]{2014AJ....147..160M}
Mann, A.~W., {Deacon}, N.~R., {Gaidos}, E., {et~al.} 2014, \aj, 147, 160,
  \dodoi{10.1088/0004-6256/147/6/160}

\bibitem[{{Mann} {et~al.}(2014){Mann}, {Deacon}, {Gaidos}, {Ansdell}, {Brewer},
  {Liu}, {Magnier}, \& {Aller}}]{Mann2014}
{Mann}, A.~W., {Deacon}, N.~R., {Gaidos}, E., {et~al.} 2014, \aj, 147, 160

\bibitem[{Mann {et~al.}(2015)Mann, {Feiden}, {Gaidos}, {Boyajian}, \& {von
  Braun}}]{2015ApJ...804...64M}
Mann, A.~W., {Feiden}, G.~A., {Gaidos}, E., {Boyajian}, T., \& {von Braun}, K.
  2015, \apj, 804, 64, \dodoi{10.1088/0004-637X/804/1/64}

\bibitem[{{Mann} {et~al.}(2019){Mann}, {Dupuy}, {Kraus}, {Gaidos}, {Ansdell},
  {Ireland}, {Rizzuto}, {Hung}, {Dittmann}, {Factor}, {Feiden}, {Martinez},
  {Ru{\'\i}z-Rodr{\'\i}guez}, \& {Chia Thao}}]{2019ApJ...871...63M}
{Mann}, A.~W., {Dupuy}, T., {Kraus}, A.~L., {et~al.} 2019, \apj, 871, 63,
  \dodoi{10.3847/1538-4357/aaf3bc}

\bibitem[{{Mawet} {et~al.}(2018){Mawet}, {Bond}, {Delorme}, {Jovanovic},
  {Cetre}, {Chun}, {Echeverri}, {Hall}, {Lilley}, {Wallace}, \&
  {Wizinowich}}]{2018SPIE10703E..06M}
{Mawet}, D., {Bond}, C.~Z., {Delorme}, J.~R., {et~al.} 2018, in Society of
  Photo-Optical Instrumentation Engineers (SPIE) Conference Series, Vol. 10703,
  Adaptive Optics Systems VI, 1070306

\bibitem[{{Metchev} {et~al.}(2015){Metchev}, {Heinze}, {Apai}, {Flateau},
  {Radigan}, {Burgasser}, {Marley}, {Artigau}, {Plavchan}, \&
  {Goldman}}]{2015ApJ...799..154M}
{Metchev}, S.~A., {Heinze}, A., {Apai}, D., {et~al.} 2015, \apj, 799, 154,
  \dodoi{10.1088/0004-637X/799/2/154}

\bibitem[{Monet {et~al.}(2003)}]{2003AJ....125..984M}
Monet, D.~G., {et~al.} 2003, \aj, 125, 984

\bibitem[{Puget {et~al.}(2004)Puget, {Stadler}, {Doyon}, {Gigan}, {Thibault},
  {Luppino}, {Barrick}, {Benedict}, {Forveille}, {Rambold}, {Thomas},
  {Vermeulen}, {Ward}, {Beuzit}, {Feautrier}, {Magnard}, {Mella}, {Preis},
  {Vallee}, {Wang}, {Lin}, {Hall}, \& {Hodapp}}]{2004SPIE.5492..978P}
Puget, P., {Stadler}, E., {Doyon}, R., {et~al.} 2004, in Society of
  Photo-Optical Instrumentation Engineers (SPIE) Conference Series, ed.
  {A.~F.~M.~Moorwood \& M.~Iye}, Vol. 5492, 978

\bibitem[{Rayner {et~al.}(2003)Rayner, {Toomey}, {Onaka}, {Denault},
  {Stahlberger}, {Vacca}, {Cushing}, \& {Wang}}]{2003PASP..115..362R}
Rayner, J.~T., {Toomey}, D.~W., {Onaka}, P.~M., {et~al.} 2003, \pasp, 115, 362

\bibitem[{Robin {et~al.}(2003)Robin, {Reyl{\'e}}, {Derri{\`e}re}, \&
  {Picaud}}]{2003A&A...409..523R}
Robin, A.~C., {Reyl{\'e}}, C., {Derri{\`e}re}, S., \& {Picaud}, S. 2003, \aap,
  409, 523, \dodoi{10.1051/0004-6361:20031117}

\bibitem[{{Rojas-Ayala} {et~al.}(2012){Rojas-Ayala}, {Covey}, {Muirhead}, \&
  {Lloyd}}]{Rojas-Ayala:2012uq}
{Rojas-Ayala}, B., {Covey}, K.~R., {Muirhead}, P.~S., \& {Lloyd}, J.~P. 2012,
  \apj, 748, 93

\bibitem[{{Saumon} \& {Marley}(2008)}]{2008ApJ...689.1327S}
{Saumon}, D., \& {Marley}, M.~S. 2008, \apj, 689, 1327, \dodoi{10.1086/592734}

\bibitem[{Scholz(2014)}]{2014A&A...561A.113S}
Scholz, R.~D. 2014, \aap, 561, A113, \dodoi{10.1051/0004-6361/201323015}

\bibitem[{{Service} {et~al.}(2016){Service}, {Lu}, {Campbell}, {Sitarski},
  {Ghez}, \& {Anderson}}]{2016PASP..128i5004S}
{Service}, M., {Lu}, J.~R., {Campbell}, R., {et~al.} 2016, \pasp, 128, 095004,
  \dodoi{10.1088/1538-3873/128/967/095004}

\bibitem[{Simons \& {Tokunaga}(2002)}]{2002PASP..114..169S}
Simons, D.~A., \& {Tokunaga}, A. 2002, \pasp, 114, 169, \dodoi{10.1086/338544}

\bibitem[{{Skrutskie} {et~al.}(2006){Skrutskie}, {Cutri}, {Stiening},
  {Weinberg}, {Schneider}, {Carpenter}, {Beichman}, {Capps}, {Chester},
  {Elias}, {Huchra}, {Liebert}, {Lonsdale}, {Monet}, {Price}, {Seitzer},
  {Jarrett}, {Kirkpatrick}, {Gizis}, {Howard}, {Evans}, {Fowler}, {Fullmer},
  {Hurt}, {Light}, {Kopan}, {Marsh}, {McCallon}, {Tam}, {Van Dyk}, \&
  {Wheelock}}]{Skrutskie2006}
{Skrutskie}, M.~F., {Cutri}, R.~M., {Stiening}, R., {et~al.} 2006, \aj, 131,
  1163

\bibitem[{{Smoluchowski} \& {Torbett}(1984)}]{1984Natur.311...38S}
{Smoluchowski}, R., \& {Torbett}, M. 1984, \nat, 311, 38,
  \dodoi{10.1038/311038a0}

\bibitem[{Tokunaga {et~al.}(2002)Tokunaga, {Simons}, \&
  {Vacca}}]{2002PASP..114..180T}
Tokunaga, A.~T., {Simons}, D.~A., \& {Vacca}, W.~D. 2002, \pasp, 114, 180,
  \dodoi{10.1086/338545}

\bibitem[{van Dam {et~al.}(2006)van Dam, {Bouchez}, {Le Mignant}, {Johansson},
  {Wizinowich}, {Campbell}, {Chin}, {Hartman}, {Lafon}, {Stomski}, \&
  {Summers}}]{2006PASP..118..310V}
van Dam, M.~A., {Bouchez}, A.~H., {Le Mignant}, D., {et~al.} 2006, \pasp, 118,
  310, \dodoi{10.1086/499498}

\bibitem[{{Vigan} {et~al.}(2016){Vigan}, {Bonnefoy}, {Ginski}, {Beust},
  {Galicher}, {Janson}, {Baudino}, {Buenzli}, {Hagelberg}, {D'Orazi},
  {Desidera}, {Maire}, {Gratton}, {Sauvage}, {Chauvin}, {Thalmann}, {Malo},
  {Salter}, {Zurlo}, {Antichi}, {Baruffolo}, {Baudoz}, {Blanchard},
  {Boccaletti}, {Beuzit}, {Carle}, {Claudi}, {Costille}, {Delboulb{\'e}},
  {Dohlen}, {Dominik}, {Feldt}, {Fusco}, {Gluck}, {Girard}, {Giro}, {Gry},
  {Henning}, {Hubin}, {Hugot}, {Jaquet}, {Kasper}, {Lagrange}, {Langlois}, {Le
  Mignant}, {Llored}, {Madec}, {Martinez}, {Mawet}, {Mesa}, {Milli},
  {Mouillet}, {Moulin}, {Moutou}, {Orign{\'e}}, {Pavlov}, {Perret}, {Petit},
  {Pragt}, {Puget}, {Rabou}, {Rochat}, {Roelfsema}, {Salasnich}, {Schmid},
  {Sevin}, {Siebenmorgen}, {Smette}, {Stadler}, {Suarez}, {Turatto}, {Udry},
  {Vakili}, {Wahhaj}, {Weber}, \& {Wildi}}]{2016AA...587A..55V}
{Vigan}, A., {Bonnefoy}, M., {Ginski}, C., {et~al.} 2016, \aap, 587, A55,
  \dodoi{10.1051/0004-6361/201526465}

\bibitem[{Wizinowich {et~al.}(2006)Wizinowich, {Le Mignant}, {Bouchez},
  {Campbell}, {Chin}, {Contos}, {van Dam}, {Hartman}, {Johansson}, {Lafon},
  {Lewis}, {Stomski}, {Summers}, {Brown}, {Danforth}, {Max}, \&
  {Pennington}}]{2006PASP..118..297W}
Wizinowich, P.~L., {Le Mignant}, D., {Bouchez}, A.~H., {et~al.} 2006, \pasp,
  118, 297, \dodoi{10.1086/499290}

\bibitem[{{Wright} {et~al.}(2010){Wright}, {Eisenhardt}, {Mainzer}, {Ressler},
  {Cutri}, {Jarrett}, {Kirkpatrick}, {Padgett}, {McMillan}, {Skrutskie},
  {Stanford}, {Cohen}, {Walker}, {Mather}, {Leisawitz}, {Gautier}, {McLean},
  {Benford}, {Lonsdale}, {Blain}, {Mendez}, {Irace}, {Duval}, {Liu}, {Royer},
  {Heinrichsen}, {Howard}, {Shannon}, {Kendall}, {Walsh}, {Larsen}, {Cardon},
  {Schick}, {Schwalm}, {Abid}, {Fabinsky}, {Naes}, \& {Tsai}}]{Wright2010}
{Wright}, E.~L., {Eisenhardt}, P.~R.~M., {Mainzer}, A.~K., {et~al.} 2010, \aj,
  140, 1868

\bibitem[{Yelda {et~al.}(2010)Yelda, {Lu}, {Ghez}, {Clarkson}, {Anderson},
  {Do}, \& {Matthews}}]{2010ApJ...725..331Y}
Yelda, S., {Lu}, J.~R., {Ghez}, A.~M., {et~al.} 2010, \apj, 725, 331,
  \dodoi{10.1088/0004-637X/725/1/331}

\bibitem[{{Zhang} {et~al.}(2019){Zhang}, {Burgasser}, {G{\'a}lvez-Ortiz},
  {Lodieu}, {Zapatero Osorio}, {Pinfield}, \& {Allard}}]{2019MNRAS.486.1260Z}
{Zhang}, Z.~H., {Burgasser}, A.~J., {G{\'a}lvez-Ortiz}, M.~C., {et~al.} 2019,
  \mnras, 486, 1260, \dodoi{10.1093/mnras/stz777}

\end{thebibliography}

\end{document}